\documentclass[aps,prb,twocolumn,amsmath,amsfonts,amssymb,floatfix,superscriptaddress]{revtex4-2}
\usepackage{epsfig,subfigure}
\usepackage{bm}
\usepackage{color,soul,xcolor}
\usepackage{hyperref}
\usepackage{wrapfig}
\usepackage{lipsum}
\hypersetup{colorlinks=false,linkcolor=black}
\usepackage[left]{lineno}

\begin{document}
\title{Two-magnon response from light scattering in altermagnets}

\author{Shuyi Li}
\email{lis3@ufl.edu}
\affiliation{Department of Physics, University of Florida, Gainesville, Florida 32611, USA}

\author{Lexu Zhao}
\affiliation{Department of Physics, University of Florida, Gainesville, Florida 32611, USA}

\author{Chunjing Jia}
\email{chunjing@phys.ufl.edu}
\affiliation{Department of Physics, University of Florida, Gainesville, Florida 32611, USA}

\begin{abstract}
Altermagnetism is a recently established class of magnetic order that combines fully compensated moments with momentum-dependent spin splitting, yet identifying its spectroscopic fingerprints remains an open challenge. In this work, we investigate finite-momentum two-magnon excitations in a two-dimensional $d$-wave altermagnet by calculating the two-magnon light scattering intensity in different polarization channels. Using linear spin-wave theory and further incorporating the leading $1/S$ quantum corrections, we demonstrate that the characteristic magnon splitting of altermagnets shifts the upper edge of the two-magnon continuum and the corresponding spectral features to higher energy at $\bm X=(\pi,0)$ relative to the conventional antiferromagnet, by an amount linear in the exchange anisotropy $|\delta J_2|$, while leaving the response at the Brillouin-zone center unchanged. Although magnon--magnon interactions strongly redistribute spectral weight toward lower energies, the energy scale associated
with the high-energy momentum-selective reconstruction remains robust. The interactions additionally generate a pronounced low-energy two-peak structure that has no counterpart within linear spin-wave theory. At $\bm K=(\pi/2,\pi/2)$, we show that the interacting
two-magnon resonance is twofold degenerate in the conventional antiferromagnetic phase, while a finite altermagnetic exchange anisotropy lifts this degeneracy, producing two peaks whose separation is linear in
$|\delta J_2|$. The characteristic energy scales underlying both features are intrinsic to the two-magnon sector rather than to a specific scattering operator. These findings highlight the potential of finite-momentum two-magnon spectroscopy for identifying altermagnetic
order in insulating magnets and motivate momentum-resolved resonant inelastic x-ray scattering studies of candidate altermagnetic materials.
\end{abstract}
\maketitle

\section{Introduction}
Altermagnetism is a recently established class of collinear magnetic order characterized by fully compensated magnetic moments and momentum-dependent spin splitting in the electronic band structure~\cite{AMth1,AMth2,AMth3,AMliu,AMth4,AMth5,AMth6,AMsplit1,AMsplit2}. The spin splitting originates from the interplay between crystal symmetry and magnetic order, allowing opposite-spin states to split even in the absence of net magnetization and spin--orbit coupling. Beyond the electronic structure, the same symmetry also influences collective spin excitations, leading to unconventional momentum-dependent splitting of the magnon branches~\cite{magnon1, magnon2, magnon3, magnon4, vb_am1}.

These unique electronic and magnetic properties have stimulated growing efforts to identify robust spectroscopic signatures of altermagnetic order. On the electronic side, angle-resolved photoemission spectroscopy (ARPES) has directly revealed the momentum-dependent spin splitting predicted for altermagnets, providing direct evidence for their unconventional band structure~\cite{ARPES1,ARPES2,ARPES3,ARPES4,ARPES5,ARPES6,ARPES7,ARPES8,ARPES9,ARPES10}. Besides electronic excitations, recent theoretical and experimental studies have shown that the characteristic chiral magnon splitting in altermagnets can be detected by momentum-resolved probes such as inelastic neutron scattering and resonant inelastic x-ray scattering (RIXS)~\cite{AM_RIXSth1,AM_RIXS1,AM_RIXS2,AM_INS1,AM_INS2,AM_INS3,AM_INS4,magnon3}. Nevertheless, existing spectroscopic investigations have focused predominantly on electronic band structures and single-magnon excitations, while the spectroscopic manifestations of altermagnetic magnon splitting in multi-magnon excitations remain largely unexplored despite a few recent theoretical studies~\cite{Multimagnon, Liu, AM_INS4}.

Among the various multi-magnon processes, two-magnon scattering has long served as a powerful probe of magnetic correlations and microscopic exchange interactions in quantum magnets~\cite{Raman-London,RIXS-TOM,RIXS-TOM2}. At zero momentum, magnetic Raman scattering probes two-magnon excitations with zero total momentum and has been extensively studied in conventional antiferromagnets~\cite{Raman-London,Raman2,Raman3,Raman4}. The corresponding effective light-scattering formalism has also been extended to finite momentum, allowing the momentum-dependent two-magnon response to be investigated throughout the Brillouin zone~\cite{RIXS-TOM,RIXS-TOM2,FQ1,FQ2,FQ3}. Such two-magnon response can also be measured by RIXS at the transition-metal $K$-edge~\cite{RIXS-Kedge-Jia, Ishii-two-magnon-Kedge, Nagao-two-magnon-Kedge}, and RIXS at the transition-metal $L$-edge with incoming and outgoing light polarizations parallel with each other~\cite{RIXS-Ledge-Jia, RIXS-two-magnon-Ament}. More generally, the resulting spectra are governed by an interplay between the two-magnon density of states (DOS), polarization-dependent scattering matrix elements, and magnon--magnon interactions, which can strongly reconstruct the peak positions and spectral line shapes~\cite{Raman2,Raman3,MgINT1,MgINT2,MgINT3,MgINT4,MgINT5,MgINT6}. Despite the extensive understanding of two-magnon excitations in conventional magnetic systems, how the characteristic magnon splitting of altermagnets is reflected in the two-magnon excitation spectrum remains an open question.

In this work, we investigate finite-momentum two-magnon excitations in a
two-dimensional $d$-wave altermagnet within an effective finite-momentum
light-scattering formalism. We calculate the momentum- and
polarization-resolved two-magnon response using linear spin-wave theory
(LSWT) supplemented by the leading $1/S$ quantum corrections, treating
the residual magnon--magnon interaction within the ladder approximation.
We show that the characteristic magnon splitting of altermagnets produces a momentum-selective reconstruction of the two-magnon spectrum, manifested by shifts of the upper continuum edge and the spectral features compared with the conventional antiferromagnet at finite momentum, while leaving the Brillouin-zone center essentially unchanged. The corresponding
high-energy scale is renormalized by the quantum corrections but remains
robust. The magnon--magnon interaction generates a low-energy
two-peak structure that is absent within LSWT. A similar splitting near
$(\pi/2,\pi/2)$ has been found in previous numerical studies~\cite{Liu,AM_INS4}, but its symmetry origin has not been clarified.
At $\bm K=(\pi/2,\pi/2)$, we find that the interacting two-magnon
resonance is twofold degenerate in the conventional antiferromagnetic
phase. The altermagnetic exchange anisotropy $\delta J_2$ lifts this
degeneracy, producing two peaks whose separation grows linearly with
$|\delta J_2|$. The above characteristic energy scales are intrinsic to the two-magnon sector rather than to a
specific scattering operator. Our results establish finite-momentum
two-magnon spectroscopy as a promising probe of altermagnetic order in
insulating magnets and provide guidance for future momentum-resolved
spectroscopic studies, including resonant inelastic x-ray scattering.

\section{Model and Formalism}

\subsection{Model Hamiltonian}

\begin{figure}
    \centering
    \hspace*{2.0cm}
    \includegraphics[width=0.6\linewidth]{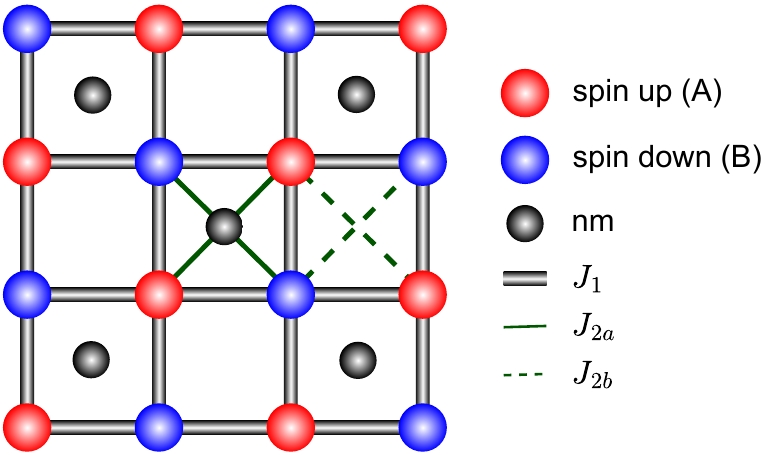}
    \caption{\textbf{Lattice structure and exchange interactions.} A schematic of the Lieb lattice and the exchange interactions in the Heisenberg model. The spin-up (A), spin-down (B), and nonmagnetic (nm) sites are shown in red, blue, and gray, respectively. The solid bonds represent the nearest-neighbor exchange interaction $J_1$, whereas the green solid and dashed bonds denote the two inequivalent next-nearest-neighbor exchange interactions $J_{2a}$ and $J_{2b}$, respectively. The limit $J_{2a}=J_{2b}$ corresponds to the conventional antiferromagnet, whereas $J_{2a}\neq J_{2b}$ gives the $d$-wave altermagnet.}
    \label{fig:Lattice}
\end{figure}

We study an effective $J_1$-$J_{2a}$-$J_{2b}$ Heisenberg model on a Lieb lattice, obtained from the corresponding half-filled $t_1$-$t_{2a}$-$t_{2b}$ Hubbard model in the strong-coupling limit $U\gg t$. The nonmagnetic sites of the Lieb lattice lower the symmetry so that the two diagonal bonds become inequivalent. The Hamiltonian is
\begin{equation}\label{HSPIN}
H=
J_1\sum_{\langle ij\rangle}\bm{S}_i\cdot\bm{S}_j
+J_{2a}\sum_{\langle\langle ij\rangle\rangle_a}\bm{S}_i\cdot\bm{S}_j
+J_{2b}\sum_{\langle\langle ij\rangle\rangle_b}\bm{S}_i\cdot\bm{S}_j,
\end{equation}
where $\bm{S}_i$ denotes the spin operator at lattice site $i$ with $S=1/2$. Here, $J_1$ is the nearest-neighbor exchange interaction, whereas $J_{2a}$ and $J_{2b}$ are the next-nearest-neighbor couplings along the two inequivalent diagonals. As illustrated in Fig.~\ref{fig:Lattice}, on sublattice A, $J_{2a}$ ($J_{2b}$) connects sites separated by $\pm\bm{d}_1=\pm(1,1)$ [$\pm\bm{d}_2=\pm(1,-1)$], in units of the lattice constant, whereas $J_{2a}$ and $J_{2b}$ are exchanged on sublattice B. For convenience, we introduce the average next-nearest-neighbor exchange interaction
\begin{equation}
J_2=\frac{J_{2a}+J_{2b}}{2},
\end{equation}
and the exchange anisotropy
\begin{equation}
\delta J_2=\frac{J_{2a}-J_{2b}}{2}.
\end{equation}
In this work, we focus on the N\'eel-ordered phase stabilized by the dominant antiferromagnetic nearest-neighbor interaction $J_1$. When the two next-nearest-neighbor couplings become inequivalent ($\delta J_2\neq0$), the system remains N\'eel ordered and enters a $d$-wave altermagnetic phase. In the following, we refer to the $\delta J_2=0$ and $\delta J_2\neq0$ cases as the conventional antiferromagnet (AFM) and the altermagnet (AM), respectively.

\subsection{Spin-Wave Theory} 
Taking the classical N\'eel-ordered state as the reference state, we perform the Holstein--Primakoff transformation on the spin operators of the two magnetic sublattices,
\begin{equation}
\begin{aligned}
    S_{Ai}^+&=(2S-a_i^{\dagger}a_i)^{\frac{1}{2}}a_i,\\ S_{Ai}^-&=a_i^{\dagger}(2S-a_i^{\dagger}a_i)^{\frac{1}{2}},\\
    S_{Ai}^z&=S-a_i^{\dagger}a_i,\\
    S_{Bj}^+&=b_j^{\dagger}(2S-b_j^{\dagger}b_j)^{\frac{1}{2}},\\
    S_{Bj}^-&=(2S-b_j^{\dagger}b_j)^{\frac{1}{2}}b_j,\\
    S_{Bj}^z&=-S+b_j^{\dagger}b_j,\\
\end{aligned}
\end{equation}
where $a_i^{\dagger}$ ($a_i$) and $b_j^{\dagger}$ ($b_j$) are the magnon creation (annihilation) operators on sublattices A and B, respectively.  Expanding the spin Hamiltonian in Eq.~(\ref{HSPIN}) in powers of $1/S$ gives
\begin{equation}
    H_M = H_0 + H_1 + H_2 + o(1/S),
\end{equation}
where $H_0$, $H_1$, and $H_2$ correspond to the classical ground-state energy, the LSW Hamiltonian, and the leading $1/S$ correction, respectively. 
After Fourier transformation, $H_1$ can be written as
\begin{equation}
    H_1=\frac{1}{2}\sum_{\bm{k}} A^{\dagger}(\bm{k}) \mathcal{H}_1(\bm{k}) A(\bm{k}),
\end{equation}
where $A(\bm{k})=(a_{\bm{k}},\, b_{\bm{k}},\, a_{-\bm{k}}^{\dagger},\, b_{-\bm{k}}^{\dagger})^{T}$, and $\mathcal{H}_1(\bm{k})$ is a $4\times4$ matrix. The Hamiltonian is diagonalized by the Bogoliubov transformation
\begin{equation}\label{BG}
a_{\bm{k}}^{\dagger}=u_{\bm{k}}\alpha_{\bm{k}}^{\dagger}+v_{\bm{k}}\beta_{-\bm{k}}, \quad
b_{-\bm{k}}=v_{\bm{k}}\alpha_{\bm{k}}^{\dagger}+
u_{\bm{k}}\beta_{-\bm{k}},
\end{equation}
where $\alpha$ and $\beta$ are the resulting magnon eigenmodes. The corresponding LSW dispersions are 
\begin{equation}
    \omega_{\pm}^{(0)}(\bm{k})=\bar{\omega}^{(0)}(\bm{k})\pm\Delta^{(0)}(\bm{k}),
\end{equation}
where $\pm$ labels the two branches with opposite chiralities, representing spin precessions with opposite rotational directions (counterclockwise and clockwise) about the ordered moment on sublattice A, as determined by the Landau--Lifshitz equation. Here, the $\bar{\omega}^{(0)}(\bm{k})$ and $\Delta^{(0)}(\bm{k})$ denote the averaged dispersion and the band splitting, respectively,
\begin{equation}
    \bar{\omega}^{(0)}(\bm{k})=\sqrt{(\frac{f_{a\bm{k}}+f_{b\bm{k}}}{2})^2-g_{\bm{k}}^2},
\end{equation}
\begin{equation}
    \Delta^{(0)}(\bm{k})=\frac{1}{2}(f_{a\bm{k}}-f_{b\bm{k}})=-4S\delta J_2\sin{k_x}\sin{k_y}.
\end{equation}
In this model, $\bar{\omega}^{(0)}(\bm{k})$ is independent of $\delta J_2$, while the band splitting $\Delta^{(0)}(\bm{k})$ is proportional to $\delta J_2$. The Bogoliubov coefficients $u_{\bm{k}}$ and $v_{\bm{k}}$, together with the matrix elements $f_{a\bm{k}}$, $f_{b\bm{k}}$, and $g_{\bm{k}}$ of $\mathcal{H}_1(\bm{k})$, are given in the Appendices. 

To incorporate quantum corrections beyond the LSW approximation, we retain the quartic term $H_2$, which gives rise to the leading $1/S$ correction to the magnon dispersions and introduces magnon--magnon interactions. Substituting the Bogoliubov transformation into $H_2$ and bringing the resulting expression into normal-ordered form gives
\begin{equation}
H_2=\mathrm{const}~+:H_1':+:H_2':,
\end{equation}
where
\begin{equation}
:H_1':=
\sum_{\bm{k}}\left[A_{\bm{k}}^+\alpha_{\bm{k}}^{\dagger}\alpha_{\bm{k}}+A_{\bm{k}}^-\beta_{\bm{k}}^{\dagger}\beta_{\bm{k}}+B_{\bm{k}}(\alpha_{\bm{k}}^{\dagger}\beta_{-\bm{k}}^{\dagger}+\alpha_{\bm{k}}\beta_{-\bm{k}})\right].
\end{equation}

The quadratic term $:H_1':$ corresponds to the Oguchi correction~\cite{Oguchi} and provides the leading $1/S$ renormalization of the magnon dispersions through the coefficients $A_{\bm{k}}^{\pm}$. The off-diagonal coefficient $B_{\bm{k}}$ contributes only at higher order in $1/S$ and is neglected here. The remaining quartic term $:H_2':$ describes the residual magnon--magnon interaction. The renormalized magnon dispersions are

\begin{equation}
\label{eq:R_energy}
\omega_{\pm}(\bm{k})=\bar{\omega}(\bm{k})\pm\Delta(\bm{k}),
\end{equation}
where
\begin{equation}
\begin{aligned}
\bar{\omega}(\bm{k})&=\bar{\omega}^{(0)}(\bm{k})+\delta\bar{\omega}(\bm{k}),\\
\Delta(\bm{k})&=\Delta^{(0)}(\bm{k})+\delta\Delta(\bm{k}).
\end{aligned}
\end{equation}
The leading corrections to the averaged dispersion and the band splitting are given by
\begin{equation}
\begin{aligned}
    \delta\bar{\omega}(\bm{k})&=J_1\frac{2-\gamma_{\bm{k}}(\cos(k_x)+\cos(k_y))}{\epsilon_{\bm{k}}}I_1\\
    &+2J_2\frac{1-\cos(k_x)\cos(k_y)}{\epsilon_{\bm{k}}}I_2,\\
\end{aligned}
\label{eq:delta1}
\end{equation}
\begin{equation}
    \delta\Delta(\bm{k})=2\delta J_2\sin(k_x)\sin(k_y)I_2,
\label{eq:delta2}
\end{equation}
so that the band splitting is renormalized by a momentum-independent
factor,
\begin{equation}
    \Delta(\bm{k})=Z_\Delta\,\Delta^{(0)}(\bm{k}),
    \qquad
    Z_\Delta=1-\frac{I_2}{2S},
    \label{eq:ZDelta}
\end{equation}
with
\begin{equation}\label{eq:I}
\begin{aligned}
    I_1&=\frac{1}{N}\sum_{\bm{q}}\frac{\gamma_{\bm{q}}\cos(q_x)+\epsilon_{\bm{q}}-1}{\epsilon_{\bm{q}}},\\
    I_2&=\frac{1}{N}\sum_{\bm{q}}\frac{1-\epsilon_{\bm{q}}-\cos(q_x)\cos(q_y)}{\epsilon_{\bm{q}}},
\end{aligned}
\end{equation}
and
\begin{equation}
\gamma_{\bm{k}}=\frac{2g_{\bm{k}}}
{f_{a\bm{k}}+f_{b\bm{k}}},~
\epsilon_{\bm{k}}=\sqrt{1-\gamma_{\bm{k}}^2}.
\end{equation}
Here $N$ is the number of magnetic unit cells. At the level of the leading $1/S$ correction, the averaged dispersion remains independent of $\delta J_2$, while the $\delta J_2$ dependence resides entirely in the band-splitting term $\Delta(\bm{k})$.

\subsection{Two-Magnon Scattering Operator}
To investigate light-induced two-magnon excitations, we consider the coupling between the electromagnetic field and the electronic degrees of freedom in the underlying Hubbard model,

\begin{equation}
H_C=-\sum_{\bm q}\bm j_{\bm q}\cdot\bm A_{-\bm q},
\end{equation}
where $\bm j_{\bm q}$ is the current operator and $\bm A_{\bm q}$ is the vector potential. Following the momentum-dependent light scattering formalism developed in Refs.~\cite{FQ1,FQ2}, we obtain the effective scattering operator for the present $t_1$-$t_{2a}$-$t_{2b}$ Hubbard model. At half filling and in the strong-coupling limit $U\gg t$, the resulting operator is
\begin{equation}\label{eq:Oq_S}
\hat{O}_{\bm{q}}=-\frac{1}{2}\sum_{ ij}(\bm{r}_{ij}\cdot\hat{\bm e}_{\text{in}})(\bm{r}_{ij}\cdot\hat{\bm e}_{\text{out}})J_{ij}\bm{S}_i\cdot\bm{S}_j e^{i\bm{q}\cdot(\bm{r}_i+\bm{r}_j)/2},
\end{equation}
where $\bm r_i$ denotes the position of site $i$,
$\bm r_{ij}=\bm r_j-\bm r_i$ is the bond vector connecting sites $i$ and $j$, $\hat{\bm e}_{\text{in}}$ and $\hat{\bm e}_{\text{out}}$ are the polarization vectors of the incoming and outgoing photons, and $\bm{q}$ is the transferred momentum. In the Raman limit $\bm q=0$, Eq.~(\ref{eq:Oq_S}) reduces to the conventional Fleury--Loudon scattering operator. Details of the derivation and convention are presented in the Appendices.

The two-magnon contribution to the scattering operator becomes
\begin{equation}
\label{eq:Oq}
    \hat{O}_{\bm{q}}=\sum_{\bm{k}}\left[ M_{\bm{k}}(\bm{q})\alpha_{\bm{k}+\bm{q}}^{\dagger}\beta_{-\bm{k}}^{\dagger}+M_{\bm{k}}^*(\bm{q})\alpha_{\bm{k}+\bm{q}}\beta_{-\bm{k}}\right],
\end{equation}
where $M_{\bm k}(\bm q)$ is the two-magnon scattering matrix element (see Appendices). Note that $\hat{O}_{\bm{q}}$ creates one magnon from each chirality branch, rather than two magnons from the same branch.

\subsection{Two-Magnon Scattering Intensity}

The quantity calculated in this work is the momentum-resolved two-magnon scattering intensity,
\begin{equation}
I(\bm{q},\omega)=-\frac{1}{\pi}\operatorname{Im}G(\bm{q},\omega),
\end{equation}
where
\begin{equation}
G(\bm{q},\omega)=-i\int_0^{\infty} dt\,
e^{i\omega t}\langle0|\mathcal{T}\hat O_{\bm q}^{\dagger}(t)\hat O_{\bm q}(0)|0\rangle
\end{equation}
is the corresponding time-ordered correlation function. 

Within LSWT, the bare correlation function is 
\begin{equation}\label{eq:G0}
G_0(\bm{q},\omega)=\frac{1}{N}\sum_{\bm{k}}|M_{\bm{k}}(\bm{q})|^2\Pi_0(\bm{q},\omega;\bm{k}),
\end{equation}
where $\Pi_0(\bm{q},\omega;\bm{k})$ is the bare two-magnon propagator
\begin{equation}
\label{eq:lsw2prop}
\Pi_0(\bm{q},\omega;\bm{k})=\frac{1}{\omega-E_2^{(0)}(\bm{k},\bm{q})+i0^+},
\end{equation}
with the two-magnon excitation energy defined as
\begin{equation}
\label{eq:E2_LSW}
E_2^{(0)}(\bm{k},\bm{q})
=\omega_+^{(0)}(\bm{k}+\bm{q})+\omega_-^{(0)}(-\bm{k}).
\end{equation}

The above expression shows that the two-magnon scattering intensity is determined jointly by the scattering matrix elements and the underlying two-magnon excitation spectrum. To isolate the latter, we introduce the corresponding two-magnon DOS,
\begin{equation}
\label{eq:lswDOS}
D(\bm{q},\omega)=\frac{1}{N}\sum_{\bm{k}}\delta\!\left[\omega
-E_2^{(0)}(\bm{k},\bm{q})\right].
\end{equation}

Beyond LSWT, we incorporate the leading $1/S$ quantum corrections by including the Oguchi correction to the one-magnon dispersions and treating the residual magnon--magnon interaction within the ladder approximation~\cite{Nagao-two-magnon-Kedge,MgINT5,Raman2}. After including the Oguchi correction given in Eqs.~(\ref{eq:delta1}) and (\ref{eq:delta2}), the renormalized single-magnon propagators are
\begin{equation}
\begin{aligned}
G_{\alpha\alpha}(\bm k,\omega)&=\frac{1}{\omega-\omega_+(\bm k)+i0^+},\\
G_{\beta\beta}(\bm k,\omega)&=\frac{1}{\omega-\omega_-(\bm k)+i0^+},\\
G_{\alpha\beta}(\bm k,\omega)&=G_{\beta\alpha}(\bm k,\omega)=0.
\end{aligned}
\end{equation}
Accordingly, the two-magnon excitation energy becomes
\begin{equation}
E_2(\bm{k},\bm{q})
=\omega_+(\bm{k}+\bm{q})+\omega_-(-\bm{k}),
\end{equation}
and the corresponding two-magnon propagator and DOS are obtained by replacing
$E_2^{(0)}$ with $E_2$ in Eqs.~(\ref{eq:lsw2prop}) and (\ref{eq:lswDOS}). In the following, $G_0$ and $\Pi_0$ denote the corresponding quantities constructed from the renormalized magnon dispersions.

The ladder approximation is implemented through the Bethe--Salpeter equation. Following the standard formalism developed in Refs.~\cite{MgINT4,Raman2}, the irreducible two-magnon interaction vertex is written in the separable form
\begin{equation}
V_{\bm k\bm k'}(\bm q)
=
\frac{1}{N}
\sum_{mn}
v_m(\bm k, \bm q)
\Gamma_{mn}(\bm q)
v_n(\bm k', \bm q),
\end{equation}
where $v_m(\bm k)$ are the basis functions and $\Gamma_{mn}(\bm q)$ is the interaction matrix in this basis. The resulting interacting correlation function is
\begin{equation}\label{eq:G}
G(\bm q,\omega)=G_0(\bm q,\omega)+G_{lad}(\bm q,\omega),
\end{equation}
where $G_{lad}(\bm q,\omega)$ represents the contribution from ladder approximation,
\begin{equation}
   G_{lad}(\bm q,\omega)=
\hat\Phi^{L}(\bm q,\omega)^{T}
\hat\Gamma(\bm q)
\left[
\hat 1
-
\hat\chi(\bm q,\omega)
\hat\Gamma(\bm q)
\right]^{-1}
\hat\Phi^{R}(\bm q,\omega), 
\end{equation}
with
\begin{equation}
\begin{aligned}
\Phi_m^{L}(\bm q,\omega)
&=
\frac{1}{N}
\sum_{\bm k}
M_{\bm k}^{*}(\bm q)
v_m(\bm k, \bm q)
\Pi_0(\bm q,\omega;\bm k),\\
\Phi_m^{R}(\bm q,\omega)
&=
\frac{1}{N}
\sum_{\bm k}
M_{\bm k}(\bm q)
v_m(\bm k, \bm q)
\Pi_0(\bm q,\omega;\bm k),\\
\chi_{mn}(\bm q,\omega)
&=
\frac{1}{N}
\sum_{\bm k}
v_m(\bm k, \bm q)
v_n(\bm k, \bm q)
\Pi_0(\bm q,\omega;\bm k).
\end{aligned}
\end{equation}
See Appendices for the details and the explicit expressions.

\section{Altermagnetic Signatures in Two-Magnon Spectra within LSWT}

\subsection{Two-magnon continuum and density of states}

\begin{figure}
\centering
\hspace{-0.7cm}
\includegraphics[width=1.05\linewidth]{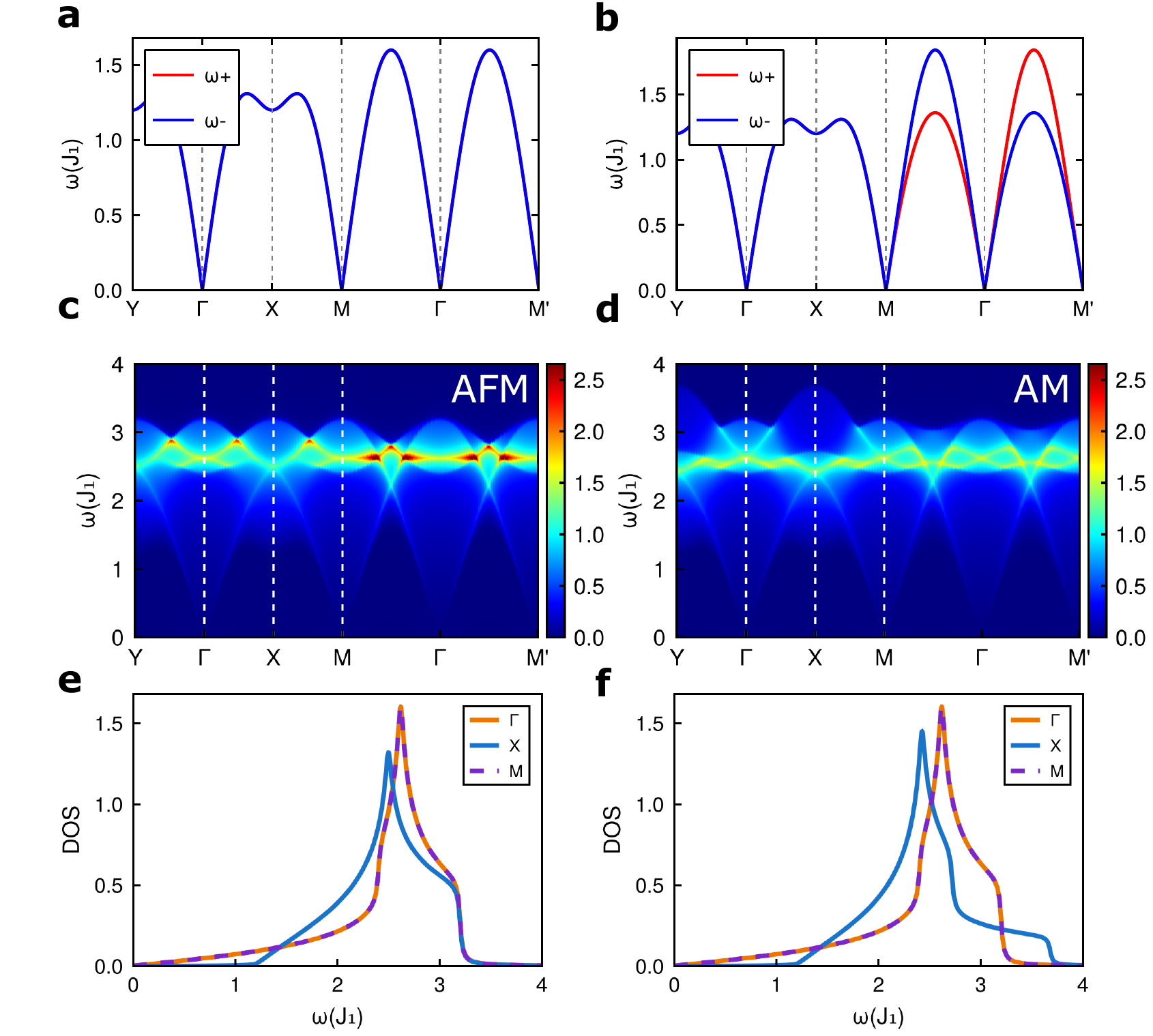}
\caption{\textbf{Magnon band structure and two-magnon DOS within LSWT.} \textbf{a} and \textbf{b} LSW magnon dispersion $\omega_\pm^{(0)}(\bm{k})$ along high-symmetry paths for the conventional antiferromagnet ($\delta J_2 = 0$) and the altermagnet ($\delta J_2 = 0.12J_1$), respectively. The red and blue lines denote the magnon branches with different chirality; they are degenerate in \textbf{a}, where only the
blue line is visible. \textbf{c} and \textbf{d} Corresponding two-magnon DOS $D(\bm{q},\omega)$ for the conventional antiferromagnet and the altermagnet. All false-color plots use a common color scale ranging from zero to \(0.8\) times the global maximum intensity. \textbf{e} and \textbf{f} Constant $\bm{q}$ cuts of $D(\bm{q},\omega)$ at the $\bm \Gamma$, $\bm M$, and $\bm X$ points for the conventional antiferromagnet and the altermagnet; the $\bm \Gamma$ and $\bm M$ curves coincide.}
    \label{fig:LSW_DOS}
\end{figure}

We first investigate the two-magnon excitation spectrum of the $d$-wave altermagnetic system within LSWT. In this work, we set $J_2=0.2J_1$, use $J_1$ as the energy unit, and treat the exchange anisotropy $\delta J_2$ as a tuning parameter. The LSW magnon dispersions $\omega_{\pm}^{(0)}(\bm{k})$ and the corresponding two-magnon DOS $D(\bm{q},\omega)$ for the conventional antiferromagnet ($\delta J_2=0$) and the $d$-wave altermagnet ($\delta J_2= 0.12J_1$) are shown in Fig.~\ref{fig:LSW_DOS}. A finite exchange anisotropy $\delta J_2$ produces the characteristic momentum-dependent splitting of the single-magnon bands and consequently reconstructs the two-magnon DOS.

This reconstruction is strongly momentum selective, as illustrated by the DOS at the $\bm \Gamma=(0,0)$, $\bm M=(\pi,\pi)$, and $\bm X=(\pi,0)$ points. Figs.~\ref{fig:LSW_DOS}\textbf{e} and \ref{fig:LSW_DOS}\textbf{f} show constant-$\bm q$ cuts of the two-magnon DOS for the conventional antiferromagnet and the altermagnet. The high-energy shoulders are associated with the upper edges of the two-magnon continua, whose energies are defined as
\begin{equation}
\omega_{\bm q}^{E}
=
\max_{\bm k} E_2^{(0)}(\bm k,\bm q).
\end{equation}
In the conventional antiferromagnet, the high-energy shoulders at $\bm \Gamma$, $\bm M$, and $\bm X$ occur at nearly the same energy, $\omega\simeq3.2J_1$. Entering the altermagnetic phase leaves the DOS at $\bm \Gamma$ and $\bm M$ unchanged, whereas the $\bm X$-point DOS undergoes a pronounced reconstruction. In particular, the main peak near $\omega\simeq2.5J_1$, which originates from a saddle-point van Hove singularity of the two-magnon continuum, shifts slightly toward lower energy, while the high-energy shoulder shifts to higher energy, reaching approximately $\omega\simeq3.68J_1$. Consequently, the nearly coincident high-energy shoulders in the conventional antiferromagnet become clearly separated in the altermagnetic phase, with only the $\bm X$-point shoulder extending to significantly higher energies than those at $\bm \Gamma$ and $\bm M$.

The momentum selectivity can be understood analytically from the two-magnon excitation energy $E_2^{(0)}(\bm k,\bm q)$. At the $\bm \Gamma$, $\bm M$, and $\bm X$ points, one obtains
\begin{equation}
\begin{aligned}
    E_2^{(0)}(\bm k, \bm \Gamma)&=E_2^{(0)}(\bm k, \bm M)=2\bar{\omega}^{(0)}(\bm k),\\
    E_2^{(0)}(\bm k, \bm X)&=\bar{\omega}^{(0)}(\bm k+\bm X)+\bar{\omega}^{(0)}(\bm k)-2\Delta^{(0)}(\bm{k}).
\end{aligned}
\end{equation}
Accordingly, the upper continuum edges are
\begin{equation}
\begin{aligned}
    \omega_{\bm \Gamma}^{E}&=\omega_{\bm M}^{E}=8S(J_1-J_2),\\
    \omega_{\bm X}^{E}&=8S(J_1-J_2)+8S|\delta J_2|.
\end{aligned}
\end{equation}
At $\bm \Gamma$ and $\bm M$, the two-magnon excitation energy is independent of $\delta J_2$, resulting in identical two-magnon DOS for the conventional antiferromagnet and the altermagnet. In contrast, $E_2^{(0)}(\bm k,\bm X)$ retains an explicit dependence on $\delta J_2$, causing the $\bm X$-point upper edge to be separated from those at $\bm \Gamma$ and $\bm M$ by $8S|\delta J_2|$. Thus, within LSWT, the $\bm X$-point reconstruction can be quantified by this upper-edge separation, which scales linearly with the magnitude of the altermagnetic exchange anisotropy.

\subsection{Polarization-resolved two-magnon spectra}

\begin{figure}
    \centering
    \includegraphics[width=0.95\linewidth]{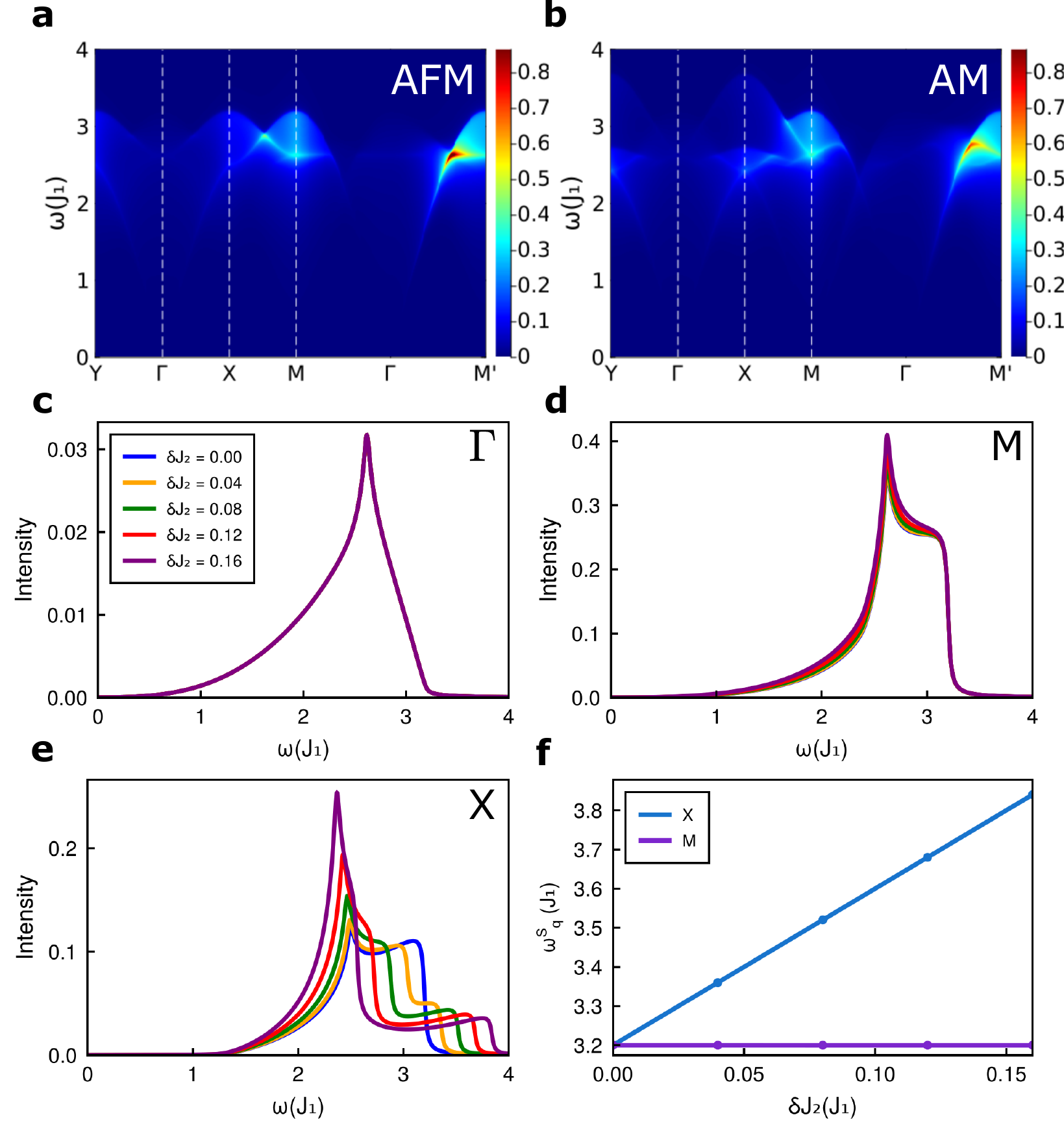}
    \caption{\textbf{Two-magnon scattering intensity in the $\hat{x}'\hat{x}'$ polarization channel within LSWT.}
    \textbf{a} and \textbf{b} Momentum-resolved intensity $I(\bm{q},\omega)$ along high-symmetry paths for the conventional antiferromagnet ($\delta J_2 = 0$) and the altermagnet ($\delta J_2 = 0.12J_1$), respectively.
    \textbf{c}, \textbf{d} and \textbf{e} Constant $\bm{q}$ cuts of the intensity at the $\bm \Gamma$, $\bm M$, and $\bm X$ points for increasing anisotropy $\delta J_2 = 0$-$0.16J_1$; the curves at $\bm\Gamma$ coincide for all $\delta J_2$.
    \textbf{f} Extracted high-energy shoulder positions $\omega_{\bm q}^{S}$ at the $\bm M$ and $\bm X$ points as functions of $\delta J_2$.}
    \label{fig:LSW_XPXP}
\end{figure}

\begin{figure}
    \centering
    \includegraphics[width=0.95\linewidth]{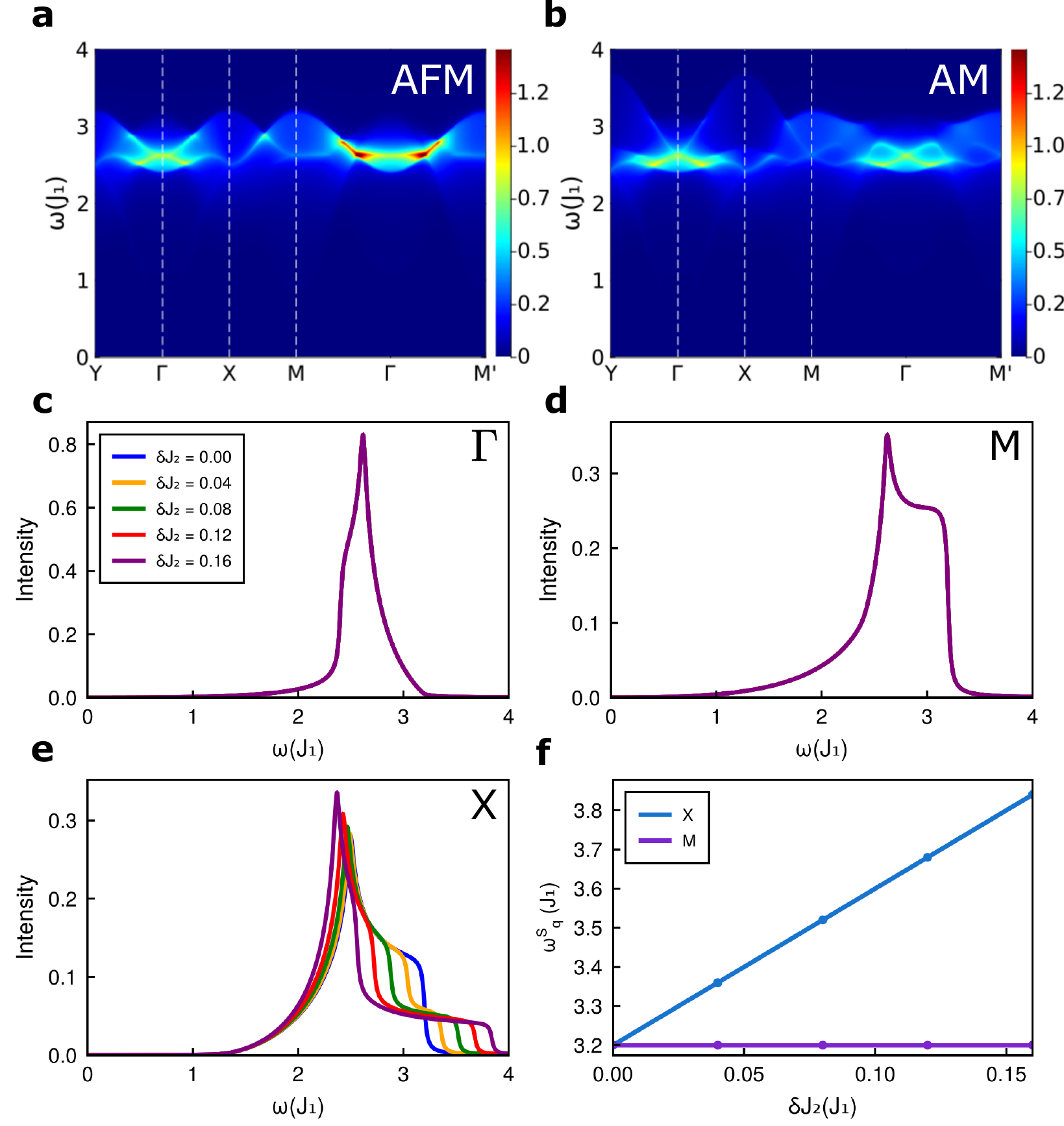}
    \caption{\textbf{Two-magnon scattering intensity in the $\hat{x}'\hat{y}'$ polarization channel within LSWT.}
    \textbf{a} and \textbf{b} Momentum--resolved intensity $I(\bm{q},\omega)$ along high-symmetry paths for the conventional antiferromagnet ($\delta J_2 = 0$) and the altermagnet ($\delta J_2 = 0.12J_1$), respectively.
    \textbf{c}, \textbf{d} and \textbf{e} Constant $\bm{q}$ cuts of the intensity at the $\bm \Gamma$, $\bm M$, and $\bm X$ points for increasing anisotropy $\delta J_2 = 0$-$0.16J_1$; the curves at $\bm\Gamma$ and $\bm M$ coincide for all $\delta J_2$.
    \textbf{f} Extracted high-energy shoulder positions $\omega_{\bm q}^{S}$ at the $\bm M$ and $\bm X$ points as functions of $\delta J_2$.}
    \label{fig:LSW_XPYP}
\end{figure}

We next turn to the polarization-resolved two-magnon scattering intensity $I(\bm{q},\omega)$ within LSWT. Figs.~\ref{fig:LSW_XPXP} and \ref{fig:LSW_XPYP} show the results in the $\hat{x}'\hat{x}'$ and $\hat{x}'\hat{y}'$ polarization channels. Although the polarization-dependent
scattering matrix elements redistribute the spectral weight, both channels exhibit the same characteristic momentum-selective reconstruction with increasing $\delta J_2$. The spectra at the $\bm \Gamma$ point are identical for the conventional antiferromagnet and the altermagnet, while those at the $\bm M$ point remain nearly unchanged. In contrast, entering the altermagnetic phase produces a pronounced reconstruction of the $\bm X$-point response. The main peak shifts slightly toward lower energy, following the displacement of the van Hove singularity in the two-magnon DOS, while the high-energy shoulder shifts continuously toward higher energy with increasing $\delta J_2$, as shown in
Figs.~\ref{fig:LSW_XPXP}\textbf{e} and \ref{fig:LSW_XPYP}\textbf{e}.

To quantify the high-energy reconstruction, we define the shoulder position $\omega_{\bm q}^{S}$ as the position of the highest-energy resolvable local maximum of $-\partial I(\bm q,\omega)/\partial\omega$. At $\bm \Gamma$, the scattering matrix element vanishes at the upper continuum edge in any polarization channel, suppressing the corresponding spectral weight.
We therefore use the $\bm M$ point as the reference for quantifying the shoulder separation, with the extracted positions summarized in Figs.~\ref{fig:LSW_XPXP}\textbf{f} and \ref{fig:LSW_XPYP}\textbf{f}. While the $\bm M$-point shoulder remains essentially unchanged, the $\bm X$-point shoulder shifts linearly toward higher energy.
Within numerical accuracy, their separation follows
\begin{equation}
    \omega_{\bm X}^{S}-\omega_{\bm M}^{S}=8S|\delta J_2|,
\end{equation}
matching the separation of the corresponding upper continuum edges in the two-magnon DOS. This agreement is not accidental, since the upper edge is determined by $\max_{\bm k}E_2^{(0)}(\bm k,\bm q)$ and therefore independent of the scattering matrix elements, which control the spectral weight near the edge.

\subsection{Circular polarization response}

\begin{figure}
    \centering
    \includegraphics[width=0.95\linewidth]{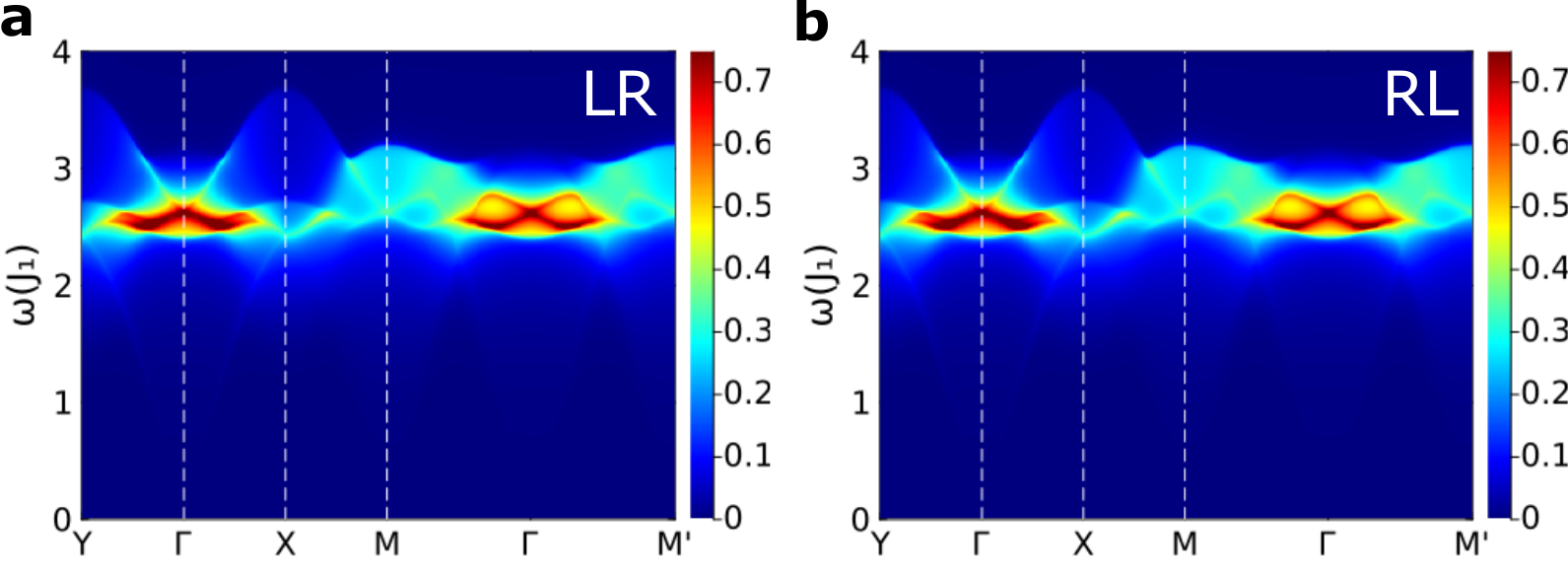}
    \caption{\textbf{Two-magnon scattering intensity in circular polarization channels within LSWT.} \textbf{a} and \textbf{b} Momentum resolved intensity $I(\bm{q},\omega)$ for the LR and RL channels of the altermagnet ($\delta J_2 = 0.12J_1$), respectively.}
    \label{fig:LSW_LRRL}
\end{figure}

Finally, we examine whether the chiral splitting of the single-magnon branches gives rise to circular dichroism in the two-magnon response. Fig.~\ref{fig:LSW_LRRL} compares the intensities in the LR and RL polarization channels, which are indistinguishable throughout the Brillouin zone. This follows directly from the structure of the scattering operator $\hat{O}_{\bm q}$ in Eq.~(\ref{eq:Oq}), where the corresponding matrix elements satisfy $M_{\bm k}^{\textit{RL}}(\bm q)=\big[M_{\bm k}^{\textit{LR}}(\bm q)\big]^{*}$, leading to the same bare correlation function. Physically, $\hat{O}_{\bm q}$ creates a pair of magnons, one from the $+$ branch and the other from the $-$ branch, rather than two magnons from the same branch, so that both circular channels probe pair states at the same energies and the chiral splitting cannot displace one channel relative to the other. Consequently, the characteristic chiral splitting of the single-magnon dispersion does not translate into circular dichroism in the two-magnon response within LSWT.

\section{Leading $1/S$ Corrections to the Two-Magnon Response}
\label{sec:1S}

We now investigate the effects of the leading $1/S$ quantum corrections on the two-magnon response. These include the Oguchi correction to the magnon dispersions and the residual magnon--magnon interaction within the ladder approximation. We begin by considering the effect of the dispersion renormalization on the two-magnon DOS and then turn to the full interacting response.

\subsection{Renormalized two-magnon continuum}

\begin{figure}
    \centering
    \hspace{-0.6cm}
    \includegraphics[width=1.05\linewidth]{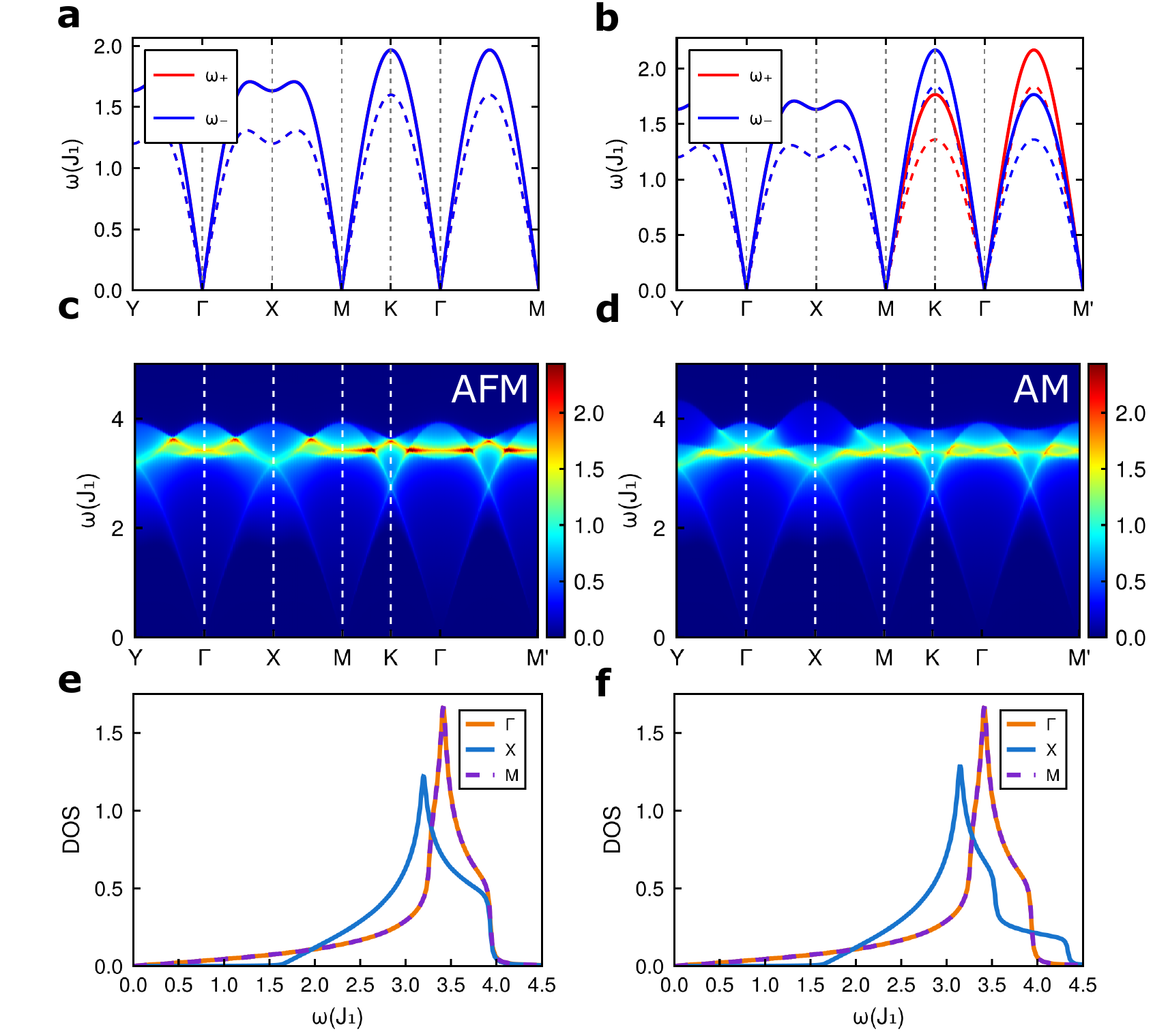}
    \caption{\textbf{$1/S$-renormalized magnon band structure and two-magnon DOS.}
    \textbf{a} and \textbf{b} Renormalized magnon dispersions $\omega_\pm(\bm{k})$ along high-symmetry paths for the conventional antiferromagnet ($\delta J_2=0$) and the altermagnet ($\delta J_2=0.12J_1$), respectively. The red and blue solid lines denote the two magnon branches, which are degenerate in \textbf{a}, while the dashed lines show the corresponding LSW dispersions $\omega_\pm^{(0)}(\bm{k})$ for comparison.
    \textbf{c} and \textbf{d} Two-magnon DOS $D(\bm{q},\omega)$ constructed from the $1/S$-renormalized magnon dispersions for the conventional antiferromagnet and the altermagnet.
    \textbf{e} and \textbf{f} Constant-$\bm q$ cuts of $D(\bm{q},\omega)$ at the $\bm \Gamma$, $\bm M$, and $\bm X$ points for the conventional antiferromagnet and the altermagnet; the $\bm\Gamma$ and $\bm M$
    curves coincide.}
    \label{fig:CR_DOS}
\end{figure}

Fig.~\ref{fig:CR_DOS} shows the $1/S$-renormalized magnon dispersions and the corresponding two-magnon DOS. Compared with LSWT, the Oguchi correction produces an overall upward renormalization of the magnon energies and shifts the two-magnon continuum toward higher energies,
while preserving the doubly degenerate bands of the conventional antiferromagnet and the characteristic $d$-wave splitting of the altermagnet. More importantly, the momentum-selective structure of the two-magnon DOS is unchanged: the $\bm \Gamma$- and $\bm M$-point DOS remain
independent of $\delta J_2$, whereas the $\bm X$-point continuum retains a pronounced dependence on the altermagnetic exchange anisotropy. Quantitatively, the separation between the upper continuum edges at $\bm X$
and $\bm \Gamma$ ($\bm M$) becomes
\begin{equation}
    \omega_{\bm X}^{E}-\omega_{\bm \Gamma}^{E}
    =\omega_{\bm X}^{E}-\omega_{\bm M}^{E}=
    8S|\delta J_2|\,Z_\Delta,
    \label{eq:edge}
\end{equation}
with $Z_\Delta$ given in Eq.~(\ref{eq:ZDelta}). For the
parameters used here, $I_2 = 0.163$ and $Z_\Delta = 0.84$, so that the $\bm X$-point reconstruction is reduced from $4\,\delta J_2$ within LSWT to $3.35\,\delta J_2$. The leading $1/S$ Oguchi correction therefore renormalizes the magnitude of the reconstruction while preserving its linear
dependence on $|\delta J_2|$. We show below that the same energy scale continues to govern the high-energy reconstruction after the residual magnon--magnon interaction is included.

\subsection{Robustness of the momentum-selective signatures}

\begin{figure}
    \centering
    \includegraphics[width=1.0\linewidth]{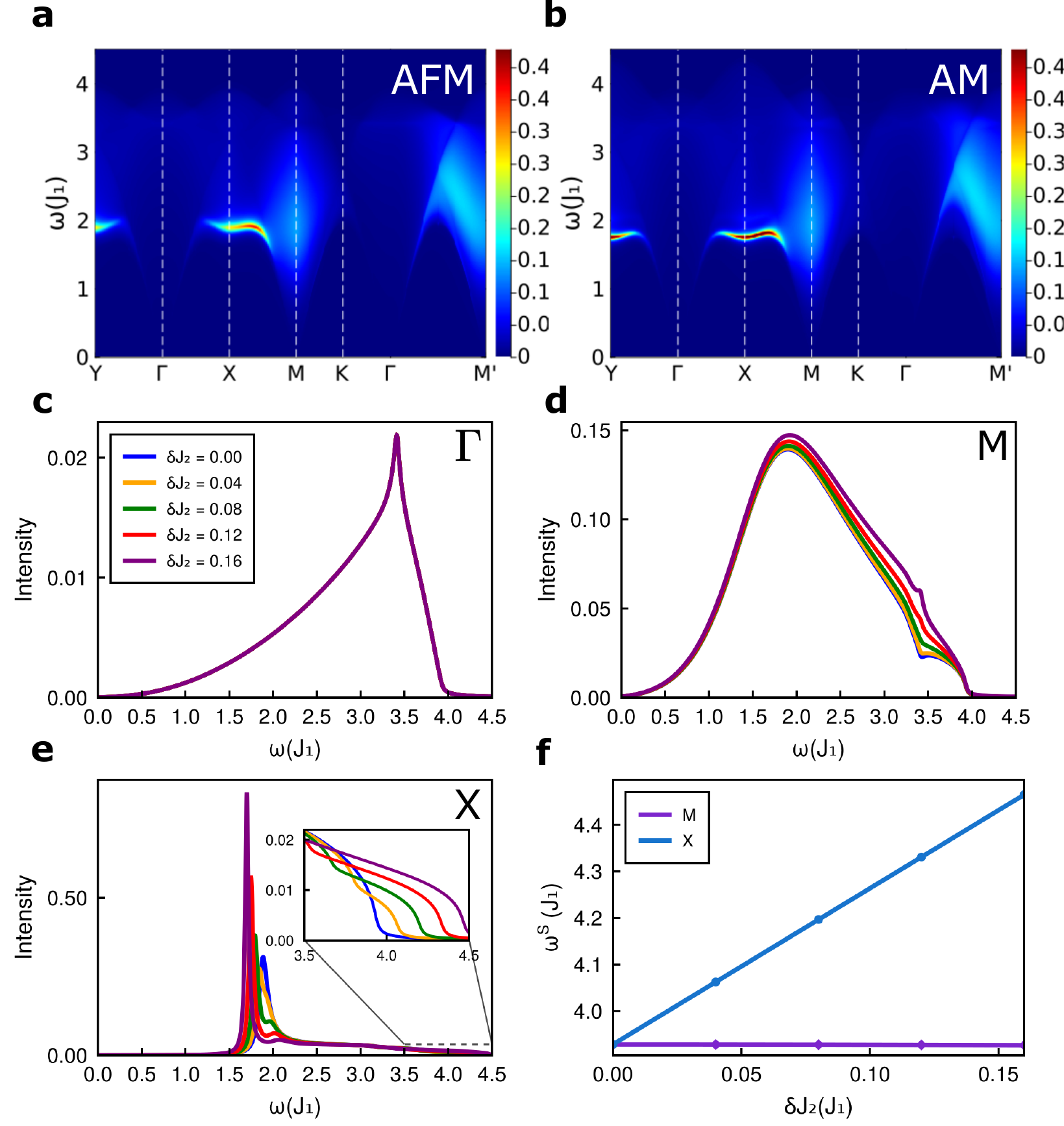}
    \caption{\textbf{Interacting two-magnon scattering intensity in the $\hat{x}'\hat{x}'$ polarization channel with leading $1/S$ corrections.}
    \textbf{a} and \textbf{b} Momentum-resolved intensity $I(\bm{q},\omega)$ along high-symmetry paths for the conventional antiferromagnet ($\delta J_2=0$) and the altermagnet ($\delta J_2=0.12J_1$), respectively.
    \textbf{c}, \textbf{d}, and \textbf{e} Constant-$\bm q$ cuts of the intensity at the $\bm \Gamma$, $\bm M$, and $\bm X$ points for increasing anisotropy $\delta J_2=0$--$0.16J_1$. The inset shows an enlargement of the high-energy region containing the shoulder.
    \textbf{f} Extracted high-energy shoulder positions
    $\omega_{\bm q}^{S}$ at the $\bm M$ and $\bm X$ points as functions of
    $\delta J_2$.}
    \label{fig:CR_XPXP}
\end{figure}

Including the residual magnon--magnon interaction within the ladder approximation reconstructs the two-magnon line shapes much more strongly than the Oguchi correction alone. As shown in Figs.~\ref{fig:CR_XPXP} and \ref{fig:CR_XPYP}, the attractive interaction transfers substantial spectral weight from the upper part of the continuum to lower energies, leaving the high-energy response comparatively weak. Nevertheless, the characteristic momentum selectivity identified within LSWT is preserved. The $\bm \Gamma$-point spectrum remains unchanged with increasing $\delta J_2$, while the $\bm M$-point response is only weakly modified. In contrast, the $\bm X$-point spectrum is strongly reconstructed, with its high-energy feature extending progressively toward larger $\omega$ as $\delta J_2$ increases, as shown in Fig.~\ref{fig:CR_XPXP}\textbf{e}.

The extracted shoulder positions shown in Fig.~\ref{fig:CR_XPXP}\textbf{f} exhibit the same dependence on $\delta J_2$ as the continuum edges in Eq.~(\ref{eq:edge}). While $\omega_{\bm M}^{S}$ is almost independent of $\delta J_2$, $\omega_{\bm X}^{S}$ grows linearly in $|\delta J_2|$ with the same slope $8S Z_\Delta$. The same slope is obtained in the $\hat{x}'\hat{y}'$ channel. This polarization independence is consistent with the high-energy shoulder tracking the underlying continuum edge, whose position is determined by $\max_{\bm k}E_2(\bm k,\bm q)$ and is therefore independent of the scattering matrix elements.

\subsection{Low-energy resonance splitting at \texorpdfstring{$\bm q=K$}{q=K}}

\begin{figure}
    \centering
    \includegraphics[width=1.0\linewidth]{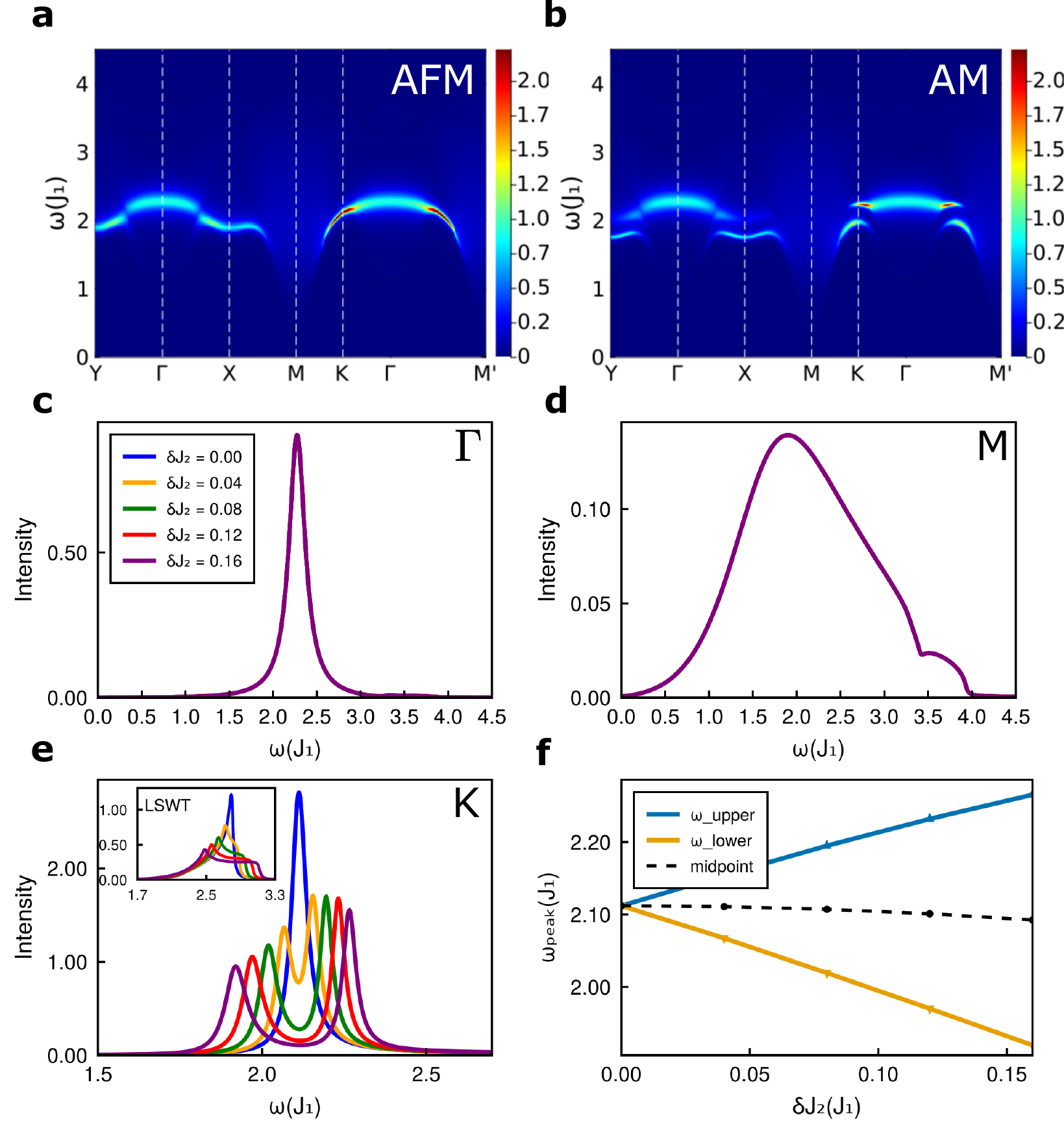}
    \caption{\textbf{Interacting two-magnon scattering intensity in the
$\hat{x}'\hat{y}'$ polarization channel with leading $1/S$ corrections.}
    \textbf{a} and \textbf{b} Momentum-resolved intensity
    $I(\bm{q},\omega)$ along high-symmetry paths for the conventional antiferromagnet ($\delta J_2=0$) and the altermagnet ($\delta J_2=0.12J_1$), respectively. \textbf{c}, \textbf{d}, and \textbf{e} Constant-$\bm q$ cuts of the intensity at the $\bm \Gamma$, $\bm M$, and $\bm K$ points for
    increasing anisotropy $\delta J_2=0$--$0.16J_1$; the curves at $\bm\Gamma$ and $\bm M$ coincide for all $\delta J_2$. Panel \textbf{e} shows an enlarged view of the low-energy resonance region at $\bm K$, while the inset shows the corresponding LSWT spectra, where this low-energy resonance structure is absent. \textbf{f} Positions of the two low-energy peaks at $\bm K$, $\omega_{\bm{K}}^{\rm upper}$ and $\omega_{\bm{K}}^{\rm lower}$, and their midpoint as functions of $\delta J_2$.}
    \label{fig:CR_XPYP}
\end{figure}

In addition to the robust high-energy momentum selectivity discussed above, a distinct interaction-induced low-energy splitting develops over a broad range of momenta, as shown for the $\hat{x}'\hat{y}'$ channel in Fig.~\ref{fig:CR_XPYP}. We focus on $\bm K=(\pi/2,\pi/2)$, where the splitting is particularly pronounced. The response exhibits a single sharp peak at $\omega_0\simeq2.11J_1$ for $\delta J_2=0$, which splits into two peaks upon introducing the altermagnetic exchange anisotropy, as shown in Fig.~\ref{fig:CR_XPYP}\textbf{e}. This low-energy two-peak structure is absent within LSWT and appears after including the magnon--magnon interaction. The evolution of the two peak positions is summarized in Fig.~\ref{fig:CR_XPYP}\textbf{f}. With increasing $|\delta J_2|$, the upper peak $\omega_{\bm{K}}^{\rm upper}$ shifts toward higher energy and the lower peak $\omega_{\bm{K}}^{\rm lower}$ toward lower energy, while their midpoint remains close to $\omega_0$. Their separation increases linearly over the whole range studied,
\begin{equation}
    \omega_{\bm{K}}^{\rm upper}-\omega_{\bm{K}}^{\rm lower}
    \simeq 2.17\,|\delta J_2|,
    \label{eq:Kdoublet}
\end{equation}
whereas the slight curvature of the individual branches is associated with a quadratic shift of the midpoint,
\begin{equation}
    \frac{\omega_{\bm{K}}^{\rm upper}+\omega_{\bm{K}}^{\rm lower}}{2}
    \simeq \omega_0-0.76\,\delta J_2^{\,2}.
    \label{eq:Kmidpoint}
\end{equation}
The behavior of the low-energy response and its splitting with increasing anisotropy $\delta J_2$ can be understood within the ladder approximation. From the interacting correlation function in Eq.~(\ref{eq:G}), the low-energy resonance is governed by the matrix
\begin{equation}
    \hat A(\bm q,\omega,\delta J_2)=\hat 1-\hat\chi(\bm q,\omega,\delta J_2)\hat\Gamma(\bm q,\delta J_2),
\end{equation}
whose near-null modes give rise to resonant enhancements in $G_{lad}$. For $\delta J_2=0$ and $\bm q=\bm K$, we denote the matrix by $\hat A_0$, which is invariant under the substitutions   
\begin{equation}
    T:\bm k\rightarrow-\bm k-\bm K,
    \qquad
    R:\bm k\rightarrow-\bm k.
\end{equation}
In the separable channel space, their matrix representations $\hat{S}_T$ and $\hat{R}_D$ satisfy
\begin{equation}
    [\hat S_T,\hat A_0]=[\hat R_D,\hat A_0]=0,~\{\hat S_T,\hat R_D\}=0,~\hat S_T^2=\hat R_D^2=\hat 1.
\end{equation}
The anticommutation is essential here. If
$\hat A_0w=a\,w$ with $\hat R_Dw=\pm w$, then
\begin{equation}
    \hat A_0(\hat S_Tw)=a\,\hat S_Tw,
    \qquad
    \hat R_D(\hat S_Tw)=\mp\hat S_Tw.
\end{equation}
Thus, $\hat S_Tw$ is a linearly independent eigenvector with the same
eigenvalue and opposite $\hat R_D$ parity. The spectrum of $\hat A_0$
is therefore twofold degenerate at every $\omega$, including the near-null resonance subspace around $\Omega_{0}$, where $\Omega_{0}$ denotes the resonance position at $\delta J_2=0$.

In the altermagnetic phase, a finite $\delta J_2$ lifts this degeneracy while preserving a symmetry relation between opposite signs of the anisotropy. At $\bm K$, the transformation $T$ exchanges the two magnons in the opposite-branch pair and reverses the sign of the altermagnetic splitting. Correspondingly, the full ladder kernel obeys
\begin{equation}
    \hat S_T
    \hat A(\bm K,\omega,\delta J_2)
    \hat S_T
    =
    \hat A(\bm K,\omega,-\delta J_2).
    \label{eq:Adelta}
\end{equation}
Meanwhile, at $\bm q=\bm K$, the $\delta J_2$-dependent terms remain even under $\hat R_D$, so that
\begin{equation}
    [\hat R_D,\hat A(\bm K,\omega,\delta J_2)]=0
\end{equation}
also for finite $\delta J_2$. The resonance components can therefore be labeled by their $\hat R_D$ parity,
\begin{equation}
    \hat R_D w_r=r w_r,
    \qquad r=\pm1.
\end{equation}
Because $\{\hat S_T,\hat R_D\}=0$, $\hat S_T$ maps the $r$ sector at $\delta J_2$ onto the $-r$ sector at $-\delta J_2$ at the same frequency. The corresponding resonance positions therefore obey
\begin{equation}
    \Omega_{\bm K,r}(\delta J_2)
    =
    \Omega_{\bm K,-r}(-\delta J_2).
    \label{eq:Kbranchsymmetry}
\end{equation}

The symmetry relation in Eq.~(\ref{eq:Kbranchsymmetry}) characterizes the intrinsic resonance structure of the interacting two-magnon kernel. Since $\hat R_D$ commutes with $\hat A$ at $\bm K$, the ladder response can be decomposed into the two $\hat R_D$ parity sectors. A near-null mode in a given sector gives rise to a resonant contribution to $G_{lad}$ whenever the external scattering vertex has a nonzero overlap with that mode. Numerically, for $\delta J_2>0$, the $r=+1$ and $r=-1$ resonance components closely track the upper and lower peaks of the full response, respectively.

Equation~(\ref{eq:Kbranchsymmetry}) further implies that the signed splitting between the two parity sectors is odd under $\delta J_2\rightarrow-\delta J_2$, whereas their midpoint is even:
\begin{equation}
\begin{aligned}
    \Omega_{\bm K,+}-\Omega_{\bm K,-}
    &=
    b_1\delta J_2+b_3\delta J_2^3
    +O(\delta J_2^5),\\
    \frac{\Omega_{\bm K,+}+\Omega_{\bm K,-}}{2}
    &=
    \Omega_0+a_2\delta J_2^2
    +O(\delta J_2^4).
\end{aligned}
\label{eq:Kexpansion}
\end{equation}
Here the subscripts $\pm$ denote the $\hat R_D=\pm1$ parity sectors. The observed peak separation and midpoint therefore closely follow the corresponding
intrinsic resonance quantities,
\begin{equation}
\begin{aligned}
    \omega_{\text{K}}^{\rm upper}
    -\omega_{\text{K}}^{\rm lower}
    &\simeq
    \left|
    \Omega_{\text{K},+}
    -\Omega_{\text{K},-}
    \right|,\\
    \frac{\omega_{\text{K}}^{\rm upper}
    +\omega_{\text{K}}^{\rm lower}}{2}
    &\simeq\frac{
    \Omega_{\text{K},+}
    +\Omega_{\text{K},-}}{2}.
\end{aligned}
\end{equation}
Consequently, the leading allowed contribution to the peak separation is linear in $|\delta J_2|$, while the midpoint has no linear correction. This is precisely the behavior reported in Eqs.~(\ref{eq:Kdoublet}) and~(\ref{eq:Kmidpoint}) and shown in Fig.~\ref{fig:CR_XPYP}\textbf{f}.

A similar splitting of the low-energy response at $(\pi/2,\pi/2)$ has been reported in tensor-network~\cite{Liu} and continuous similarity
transformation~\cite{AM_INS4} calculations of the longitudinal dynamical structure factor, which also probes two-magnon excitations. In particular, in Ref.~\cite{Liu} the peak splitting vanishes at zero anisotropy and grows approximately linearly at small anisotropy, with the two branches shifting by nearly equal and opposite amounts and their midpoint remaining almost unchanged, which is consistent with our result. Here, our ladder analysis identifies the splitting in the altermagnet as the lifting of a symmetry-enforced twofold degeneracy of the low-energy two-magnon resonance. The resonance positions are determined by the intrinsic kernel $\hat A$, rather than by the external scattering matrix elements. Different probes coupling to the same two-magnon sector can therefore reflect the same underlying resonance splitting, while the probe and polarization determine its spectral weight and visibility.

\subsection{Absence of circular dichroism}

\begin{figure}
    \centering
    \includegraphics[width=0.95\linewidth]{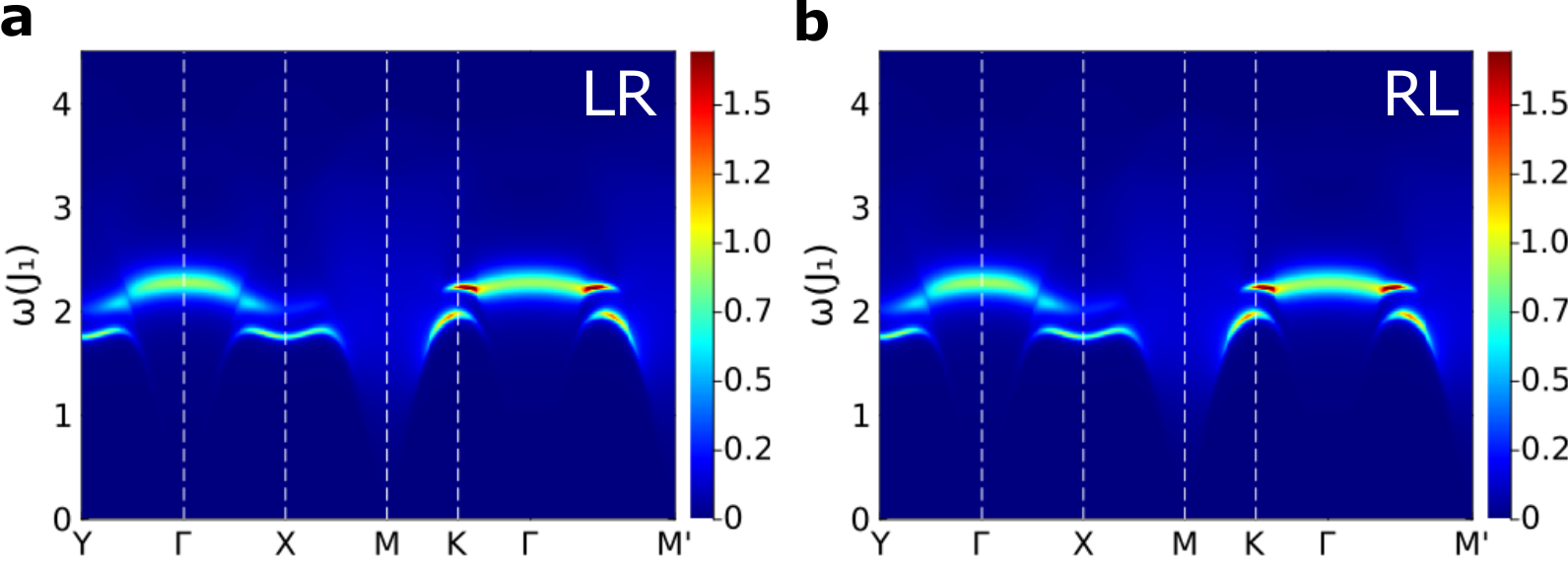}
    \caption{\textbf{Interacting two-magnon scattering intensity in circular polarization channels with leading $1/S$ corrections.}
    \textbf{a} and \textbf{b} Momentum-resolved intensity $I(\bm{q},\omega)$ for the LR and RL channels of the altermagnet ($\delta J_2=0.12J_1$), respectively.}
    \label{fig:CR_LRRL}
\end{figure}

As shown in Fig.~\ref{fig:CR_LRRL}, the LR and RL responses remain indistinguishable after including the leading $1/S$ corrections. This equality follows from the interacting correlation function in Eq.~(\ref{eq:G}). Since $M_{\bm k}^{\textit{RL}}=\big[M_{\bm k}^{\textit{LR}}\big]^*$ and the matrix $\hat\Gamma(\hat{1}-\hat\chi\hat\Gamma)^{-1}$ is symmetric, interchanging the LR and RL channels is equivalent to taking the transpose of $G(\bm q,\omega)$, which leaves the scalar $G$ unchanged. The LR and RL intensities are therefore identical to all orders within the ladder approximation.

\section{Discussion}
In this work, we investigate finite-momentum light-induced two-magnon scattering in a two-dimensional $d$-wave altermagnet.

Our LSWT results demonstrate that the two-magnon spectroscopic signatures of
altermagnetism are intrinsically momentum selective. Upon going from the antiferromagnetic
to the altermagnetic phase, the scattering intensity remains unchanged at $\bm \Gamma$ and
is only weakly modified at $\bm M$, whereas pronounced spectral reconstruction
emerges at the $\bm X$ and symmetry-related $\bm Y$ points. More generally, in
collinear two-sublattice altermagnets where the altermagnetic exchange enters only through intra-sublattice couplings, this momentum selectivity is
governed by the symmetry of the momentum-dependent magnon splitting rather
than by the specific details of the model. The splitting-dependent
contribution to the two-magnon excitation energy is
$\Delta(\bm{k+q})-\Delta(-\bm{k})$. In the present $d$-wave altermagnet,
this contribution cancels exactly at the $\bm \Gamma$ and $\bm M$ points, leaving the
two-magnon continuum unchanged. Near the $\bm X$ and $\bm Y$ points, however, the two
magnons sample regions with opposite signs of the $d$-wave splitting, so that
the two contributions add rather than cancel. As a result, the upper boundary
of the continuum shifts toward higher energies, giving rise to the
characteristic displacement of the high-energy shoulder. The same argument
naturally extends to other altermagnetic symmetries, such as $g$- and
$i$-wave spin splitting, leading to analogous finite-momentum reconstructions
with the characteristic momenta determined by the corresponding splitting
symmetry.

Going beyond LSWT, the leading $1/S$ corrections act on the two-magnon
response in three distinct ways. First, the Oguchi correction
renormalizes the altermagnetic magnon splitting by the factor
$Z_\Delta=1-I_2/2S$, which reduces the momentum-selective energy
separation from $8S|\delta J_2|$ to $8S Z_\Delta|\delta J_2|$.
Second, the residual magnon--magnon interaction transfers substantial
spectral weight from the upper part of the continuum toward lower
energies, strongly reshaping the line shapes while preserving the
underlying high-energy scale, so that the shoulder separation continues
to follow Eq.~(\ref{eq:edge}). Finally, the residual interaction
generates a qualitatively new low-energy resonance structure that is
absent within LSWT. The resulting peak splitting develops over a broad
range of momenta and is particularly pronounced at
$\bm K=(\pi/2,\pi/2)$. At this momentum, our symmetry analysis shows
that the single peak in the conventional antiferromagnetic phase
corresponds to a twofold-degenerate two-magnon resonance, whose
degeneracy is lifted by the altermagnetic exchange anisotropy. The
resulting doublet has a separation that grows linearly with
$|\delta J_2|$, while its midpoint changes only quadratically.

Importantly, these two signatures reflect intrinsic structures of the
two-magnon sector rather than specific properties of the
light-scattering operator. The upper continuum edge is determined by
$\max_{\bm k}E_2(\bm k,\bm q)$, whereas the low-energy resonance
positions at $\bm K$ are determined by the ladder kernel $\hat A$;
neither depends on the external scattering matrix elements. The probe
and polarization instead control the spectral weight and visibility of
these features. Within the present light-scattering response, this is
consistent with the high-energy shoulder separation following the same
underlying energy scale in the $\hat{x}'\hat{x}'$ and
$\hat{x}'\hat{y}'$ channels. More generally, other probes coupling to the same two-magnon sector can reveal analogous interaction-induced structures. This is consistent
with the low-energy splitting reported at $(\pi/2,\pi/2)$ in
calculations of the longitudinal dynamical structure
factor~\cite{Liu,AM_INS4}.

From the perspective of experiment, our results highlight the
importance of finite momentum for detecting the two-magnon signatures
of altermagnetism. The response at the Brillouin-zone center is
insensitive to the altermagnetic exchange anisotropy, so that
conventional Raman scattering alone cannot distinguish the conventional
antiferromagnetic and altermagnetic phases in the present setting.
At finite $\bm q$, the interaction-induced low-energy splitting at
$\bm K$ provides the most pronounced signature, while the
momentum-selective reconstruction of the high-energy continuum edge
survives as a weaker feature near the $\bm X$ and $\bm Y$ points. Our
result provides a framework for identifying altermagnetic order, and in
favorable cases for estimating the altermagnetic exchange anisotropy,
from momentum-resolved measurements of two-magnon excitations, such as
RIXS, in insulating magnets.

\section{Acknowledgments}

We thank Tom Devereaux and Brian Moritz for the insightful discussions. This work is supported by the Center for Molecular Magnetic Quantum Materials, an Energy Frontier Research Center funded by the U.S. Department of Energy, Office of Science, Basic Energy Sciences under Award no. DE-SC0019330. Computations were done using the utilities of the University of Florida Research Computing.

\appendix
\section{Linear Spin-Wave Theory}
The effective $J_1$-$J_{2a}$-$J_{2b}$ Heisenberg model considered in this work is given by
\begin{equation}\label{HSPIN-A}
H=
J_1\sum_{\langle ij\rangle}\bm{S}_i\cdot\bm{S}_j
+J_{2a}\sum_{\langle\langle ij\rangle\rangle_a}\bm{S}_i\cdot\bm{S}_j
+J_{2b}\sum_{\langle\langle ij\rangle\rangle_b}\bm{S}_i\cdot\bm{S}_j.
\end{equation}
After the Holstein--Primakoff transformation and Fourier transformation, the LSW Hamiltonian $H_1$ takes the form
\begin{equation}
    H_1=\frac{1}{2}\sum_{\bm{k}} A^{\dagger}(\bm{k}) \mathcal{H}_1(\bm{k}) A(\bm{k}),
\end{equation}
where $A(\bm{k})=(a_{\bm{k}},\, b_{\bm{k}},\, a_{-\bm{k}}^{\dagger},\, b_{-\bm{k}}^{\dagger})^{T}$, and the $4\times4$ matrix $\mathcal{H}_1(\bm{k})$ is,
\begin{equation}
H_1(\bm{k})=
\begin{bmatrix}
f_{a\bm{k}} & 0 & 0 & g_{\bm{k}}\\
0 & f_{b\bm{k}} & g_{\bm{k}} & 0\\
0 & g_{\bm{k}} & f_{a\bm{k}} & 0\\
g_{\bm{k}} & 0 & 0 & f_{b\bm{k}}\\
\end{bmatrix},
\end{equation}
with
\begin{equation}
\begin{aligned}
f_{a\bm{k}}&=4J_1S-4J_2S\\
&+2J_{2a}S\cos(k_x+k_y)+2J_{2b}S\cos(k_x-k_y),\\
f_{b\bm{k}}&=4J_1S-4J_2S\\
&+2J_{2b}S\cos(k_x+k_y)+2J_{2a}S\cos(k_x-k_y),\\
g_{\bm{k}}&=2J_1S(\cos(k_x)+\cos(k_y)).
\end{aligned}
\end{equation}
The Hamiltonian is diagonalized by the Bogoliubov transformation, 
\begin{equation}\label{BG-A}
a_{\bm{k}}^{\dagger}=u_{\bm{k}}\alpha_{\bm{k}}^{\dagger}+v_{\bm{k}}\beta_{-\bm{k}},~
b_{-\bm{k}}=v_{\bm{k}}\alpha_{\bm{k}}^{\dagger}+
u_{\bm{k}}\beta_{-\bm{k}},
\end{equation}
where
\begin{equation}
\begin{aligned}
u_{\bm{k}}&=\left(\frac{1+\epsilon_{\bm{k}}}
{2\epsilon_{\bm{k}}}\right)^{1/2},\\
v_{\bm{k}}&=-\mathrm{sgn}(\gamma_{\bm{k}})
\left(\frac{1-\epsilon_{\bm{k}}}{2\epsilon_{\bm{k}}}\right)^{1/2},\\
\epsilon_{\bm{k}}&=\sqrt{1-\gamma_{\bm{k}}^2},
\\
\gamma_{\bm{k}}&=\frac{2g_{\bm{k}}}
{f_{a\bm{k}}+f_{b\bm{k}}}.
\end{aligned}
\end{equation}

The resulting magnon eigenmodes $\alpha$ and $\beta$ have LSW dispersions
\begin{equation}
    \omega_{\pm}^{(0)}(\bm{k})=\sqrt{(\frac{f_{a\bm{k}}+f_{b\bm{k}}}{2})^2-g_{\bm{k}}^2}\pm\frac{1}{2}(f_{a\bm{k}}-f_{b\bm{k}}).
\end{equation}

\section{Effective Two-Magnon Scattering Operator}
\label{app:scattering_operator}

In this appendix, we summarize the derivation of the effective
finite-momentum light-scattering operator used in the main text, following Ref.~\cite{FQ1,FQ2}. The coupling between the electromagnetic field and the electronic
degrees of freedom of the underlying Hubbard model is
\begin{equation}
H_C=-\sum_{\bm q}\bm j_{\bm q}\cdot\bm A_{-\bm q},
\end{equation}
where $\bm j_{\bm q}$ is the current operator and
$\bm A_{\bm q}$ is the vector potential.

To second order in the light--matter coupling, the scattering amplitude between the initial and final states $|i\rangle$ and $|f\rangle$ is
\begin{equation}
\begin{aligned}
\langle f|\mathcal M|i\rangle=
\sum_{\nu}\bigg[&
\frac{
\langle f|
\bm j_{\bm k_{\mathrm{out}}}
\cdot\hat{\bm e}_{\mathrm{out}}
|\nu\rangle
\langle \nu|
\bm j_{-\bm k_{\mathrm{in}}}
\cdot\hat{\bm e}_{\mathrm{in}}
|i\rangle}{E_{\nu}-E_i-\omega_{\mathrm{in}}}
\\
+&
\frac{
\langle f|
\bm j_{-\bm k_{\mathrm{in}}}
\cdot\hat{\bm e}_{\mathrm{in}}
|\nu\rangle
\langle \nu|
\bm j_{\bm k_{\mathrm{out}}}
\cdot\hat{\bm e}_{\mathrm{out}}
|i\rangle}{E_{\nu}-E_i+\omega_{\mathrm{out}}
}
\bigg],
\end{aligned}
\label{eq:KH_amplitude}
\end{equation}
where $\hat{\bm e}_{\mathrm{in}}$ and
$\hat{\bm e}_{\mathrm{out}}$ denote the polarization vectors of the
incoming and outgoing photons, following the convention of
Ref.~\cite{FQ2,RIXS-TOM}, and $\bm q=\bm k_{\mathrm{out}}-\bm k_{\mathrm{in}}$ is the transferred momentum.

In the $t_1$-$t_{2a}$-$t_{2b}$ Hubbard model, the current operator is given by
\begin{equation}
\begin{aligned}
\bm{j}_{\bm{q}}&=\sum_{\bm{k},\alpha,\beta}\nabla_{\bm{k}}H_{\bm{k}}^{\alpha\beta}c_{\alpha\bm{k}+\bm{q}/2}^{\dagger}c_{\beta\bm{k}-\bm{q}/2}\\
&=-i\sum_{\langle ij\rangle,\sigma}t_{ij}\,
\bm r_{ij}\left(c_{i\sigma}^{\dagger}c_{j\sigma}
-c_{j\sigma}^{\dagger}c_{i\sigma}\right)
e^{i\bm q\cdot(\bm r_i+\bm r_j)/2},
\end{aligned}
\end{equation}
where
\begin{equation}
\bm r_{ij}=\bm r_j-\bm r_i
\end{equation}
is the bond vector. The phase factor reflects the finite transferred momentum of the scattering process.

In the half-filled strong-coupling limit $U\gg t$, the
initial and final states $|i\rangle$ and $|f\rangle$ belong to the singly occupied subspace, whereas
the intermediate states $|\nu\rangle$ contain one doublon--holon pair. Approximating their excitation energies by
\begin{equation}
E_{\nu}-E_i\simeq U,
\end{equation}
the sum over intermediate states can be replaced by a projection onto the subspace containing one doublon--holon pair. At order $t^2/U$, the second hopping process must annihilate the doublon--holon pair created by the first, thereby returning the system to the singly occupied subspace. Consequently, only hopping processes on the same bond contribute.

Finally, using the operator identity in the singly occupied subspace 
\begin{equation}
\sum_{\sigma\sigma'}
c_{i\sigma}^{\dagger}c_{j\sigma}
c_{j\sigma'}^{\dagger}c_{i\sigma'}=
\frac{1}{2}-2\bm S_i\cdot\bm S_j,
\label{eq:spin_projection}
\end{equation}
and omitting the constant term, we obtain the effective finite-momentum scattering operator
\begin{equation}
\hat O_{\bm q}=\mathcal C
\sum_{\langle ij\rangle}
\left(
\bm r_{ij}\cdot\hat{\bm e}_{\mathrm{in}}
\right)
\left(
\bm r_{ij}\cdot\hat{\bm e}_{\mathrm{out}}
\right)
J_{ij}\,
\bm S_i\cdot\bm S_j\,
e^{i\bm q\cdot(\bm r_i+\bm r_j)/2},
\label{eq:effective_scattering_operator-A}
\end{equation}
where $J_{ij}=4t_{ij}^{\,2}/U$, and $\mathcal C$ contains the photon-energy denominators and overall constants that are independent of the spin and momentum structure.

For the present $t_1$-$t_{2a}$-$t_{2b}$ model, $J_{ij}$ takes the values $J_1$, $J_{2a}$, and $J_{2b}$ on the corresponding bonds shown in Fig.~\ref{fig:Lattice}. In the Raman limit $\bm q=0$, Eq.~(\ref{eq:effective_scattering_operator-A}) reduces to the conventional Fleury--Loudon operator,
\begin{equation}
\hat O_{\bm q=\bm 0}
=
\mathcal C
\sum_{\langle ij\rangle}
\left(
\bm r_{ij}\cdot\hat{\bm e}_{\mathrm{in}}
\right)
\left(
\bm r_{ij}\cdot\hat{\bm e}_{\mathrm{out}}
\right)
J_{ij}\,
\bm S_i\cdot\bm S_j.
\end{equation}

Upon substituting the Bogoliubov transformation in Eq.~(\ref{BG-A}) into
Eq.~(\ref{eq:effective_scattering_operator-A}), the two-magnon contribution
to the scattering operator takes the form
\begin{equation}
\label{eq:Oq_LSW}
\hat{O}_{\bm{q}}
=
\sum_{\bm{k}}
\left[
M_{\bm{k}}(\bm{q})
\alpha_{\bm{k}+\bm{q}}^{\dagger}
\beta_{-\bm{k}}^{\dagger}
+
M_{\bm{k}}^*(\bm{q})
\alpha_{\bm{k}+\bm{q}}
\beta_{-\bm{k}}
\right],
\end{equation}
where
\begin{equation}
\begin{aligned}
M_{\bm{k}}(\bm{q})
={}&
\left[
F_1(\bm{q})+F_A(\bm{k},\bm{q})
\right]
u_{\bm{k}+\bm{q}}v_{\bm{k}}
\\
&+
\left[
F_1(\bm{q})+F_B(\bm{k},\bm{q})
\right]
u_{\bm{k}}v_{\bm{k}+\bm{q}}
\\
&+
F_2(\bm{k},\bm{q})
\left(
u_{\bm{k}+\bm{q}}u_{\bm{k}}
+
v_{\bm{k}+\bm{q}}v_{\bm{k}}
\right),
\end{aligned}
\label{eq:Mkq}
\end{equation}
with
\begin{equation}
F_1(\bm{q})
=
-2J_1S
\left(
P_x\cos\frac{q_x}{2}
+
P_y\cos\frac{q_y}{2}
\right),
\end{equation}
\begin{equation}
\begin{aligned}
F_A(\bm{k},\bm{q})
={}&
-2J_{2a}SP_{xy}^{(1)}C_{+}(\bm{k},\bm{q})
-2J_{2b}SP_{xy}^{(2)}C_{-}(\bm{k},\bm{q}),
\\
F_B(\bm{k},\bm{q})
={}&
-2J_{2b}SP_{xy}^{(1)}C_{+}(\bm{k},\bm{q})
-2J_{2a}SP_{xy}^{(2)}C_{-}(\bm{k},\bm{q}),
\end{aligned}
\end{equation}
\begin{equation}
F_2(\bm{k},\bm{q})
=
-2J_1S
\left[
P_x\cos\left(k_x+\frac{q_x}{2}\right)
+
P_y\cos\left(k_y+\frac{q_y}{2}\right)
\right],
\end{equation}
and
\begin{align}
C_{+}(\bm{k},\bm{q})
&=
-\cos\frac{q_x+q_y}{2}
+
\cos\left(
k_x+k_y+\frac{q_x+q_y}{2}
\right),
\\
C_{-}(\bm{k},\bm{q})
&=
-\cos\frac{q_x-q_y}{2}
+
\cos\left(
k_x-k_y+\frac{q_x-q_y}{2}
\right).
\end{align}

Here, $P_x$, $P_y$, $P_{xy}^{(1)}$, and $P_{xy}^{(2)}$ are determined by
the polarization vectors of the incoming and outgoing photons. Their values
for the polarization channels considered in this work are listed in
Table~\ref{table:polarization_factors}.

\begin{table}
\centering
\begin{tabular}{|c|c|c|c|c|c|}
\hline
In & Out & $P_x$ & $P_y$ & $P_{xy}^{(1)}$ & $P_{xy}^{(2)}$ \\
\hline
$x$  & $x$  & $1$   & $0$    & $1$  & $1$  \\
$y$  & $y$  & $0$   & $1$    & $1$  & $1$  \\
$x$  & $y$  & $0$   & $0$    & $1$  & $-1$ \\
$y$  & $x$  & $0$   & $0$    & $1$  & $-1$ \\
$x'$ & $x'$ & $1/2$ & $1/2$  & $2$  & $0$  \\
$y'$ & $y'$ & $1/2$ & $1/2$  & $0$  & $2$  \\
$x'$ & $y'$ & $1/2$ & $-1/2$ & $0$  & $0$  \\
$y'$ & $x'$ & $1/2$ & $-1/2$ & $0$  & $0$  \\
$L$  & $R$  & $1/2$ & $-1/2$ & $i$  & $-i$ \\
$R$  & $L$  & $1/2$ & $-1/2$ & $-i$ & $i$  \\
$L$  & $L$  & $1/2$ & $1/2$  & $1$  & $1$  \\
$R$  & $R$  & $1/2$ & $1/2$  & $1$  & $1$  \\
\hline
\end{tabular}
\caption{Polarization factors entering the two-magnon scattering matrix
element $M_{\bm{k}}(\bm{q})$ for different incoming and outgoing photon
polarizations.}
\label{table:polarization_factors}
\end{table}

\section{Interacting Two-Magnon Response in the Ladder Approximation}
\label{app:ladder}

In this appendix, we outline the solution of the Bethe--Salpeter equation for the interacting two-magnon response within the ladder approximation, following the approach of Ref.~\cite{MgINT4,Raman2}. The momentum-resolved two-magnon scattering intensity is given by
\begin{equation}
I(\bm{q},\omega)=-\frac{1}{\pi}\operatorname{Im}G(\bm{q},\omega),
\end{equation}
where
\begin{equation}
\begin{aligned}
G(\bm{q},\omega)&=-i\int_0^{\infty} dt\,
e^{i\omega t}\langle0|\mathcal{T}\hat O_{\bm q}^{\dagger}(t)\hat O_{\bm q}(0)|0\rangle\\
&=\frac{1}{N}\sum_{\bm{k}\bm{k}'}M^*_{\bm{k}}(\bm{q})M_{\bm{k}'}(\bm{q})\Pi(\bm{q},\omega;\bm{k},\bm{k}'),
\end{aligned}
\label{eq:G_app}
\end{equation}
and the two-magnon propagator $\Pi$ can be expanded as
\begin{equation}
\begin{aligned}
\Pi(\bm q,\omega;\bm k,\bm k')
=&
i\int\frac{d\omega_1}{2\pi}\,
G_{\alpha\alpha}(\bm k+\bm q,\omega+\omega_1)\\
&\times
G_{\beta\beta}(-\bm k,-\omega_1)
\Lambda_{\bm k\bm k'}(\bm q,\omega,\omega_1),
\end{aligned}
\label{eq:Pi_vertex}
\end{equation}
with the single--magnon propagators to $1/S$ order
\begin{equation}
\begin{aligned}
G_{\alpha\alpha}(\bm k,\omega)&=\frac{1}{\omega-\omega_+(\bm k)+i0^+},\\
G_{\beta\beta}(\bm k,\omega)&=\frac{1}{\omega-\omega_-(\bm k)+i0^+},
\end{aligned}
\end{equation}
and the vertex function $\Lambda$ satisfies the Bethe--Salpeter equation
\begin{equation}
\begin{aligned}
\Lambda_{\bm k\bm k'}(\bm q,\omega,\omega')
=&
\delta_{\bm k\bm k'}
+
i\sum_{\bm k_1}
\int\frac{d\omega_1}{2\pi}\,
V_{\bm k\bm k_1}(\bm q)\Lambda_{\bm k_1\bm k'}
(\bm q,\omega,\omega_1)\,\\
&\times G_{\alpha\alpha}
(\bm k_1+\bm q,\omega+\omega_1)
G_{\beta\beta}(-\bm k_1,-\omega_1),
\end{aligned}
\label{eq:BS_full_vertex}
\end{equation}
where the irreducible two-magnon interaction vertex $V$ can be written in the separable form
\begin{equation}
V_{\bm k\bm k'}(\bm q)
=
\frac{1}{N}
\sum_{m,n}
v_m(\bm k, \bm q)\,
\Gamma_{mn}(\bm q)\,
v_n(\bm k',\bm q).
\label{eq:separable_vertex_app}
\end{equation}
After substituting Eq.~(\ref{eq:Pi_vertex}) and (\ref{eq:BS_full_vertex}) into Eq.~(\ref{eq:G_app}), the correlation function becomes
\begin{equation}
    G(\bm q,\omega)=G_0(\bm q,\omega)+G_{lad}(\bm q,\omega),
\end{equation}
where
\begin{equation}
G_0(\bm q,\omega)
=
\frac{1}{N}
\sum_{\bm k}
|M_{\bm k}(\bm q)|^2
\Pi_0(\bm q,\omega;\bm k)
\label{eq:G0_app}
\end{equation}
is the noninteracting correlation function, $\Pi_0(\bm q,\omega;\bm k)$ is the non-interacting two-magnon propagator
\begin{equation}
\Pi_0(\bm{q},\omega;\bm{k})=\frac{1}{\omega-\omega_+(\bm{k}+\bm{q})-\omega_-(-\bm{k})+i0^+}.
\end{equation}
The ladder contribution is
\begin{equation}
G_{lad}(\bm q,\omega)
=
(\hat\Phi^L(\bm q,\omega))^{T}
\hat\Gamma(\bm q)
\hat \Psi^R(\bm q,\omega),
\label{eq:GL_B}
\end{equation}
where
\begin{equation}
\begin{aligned}
\Phi_m^{L}(\bm q,\omega)
&=
\frac{1}{N}
\sum_{\bm k}
M_{\bm k}^{*}(\bm q)
v_m(\bm k, \bm q)
\Pi_0(\bm q,\omega;\bm k),\\
\Phi_m^{R}(\bm q,\omega)
&=
\frac{1}{N}
\sum_{\bm k}
M_{\bm k}(\bm q)
v_m(\bm k, \bm q)
\Pi_0(\bm q,\omega;\bm k),
\end{aligned}
\end{equation}
and
\begin{equation}
\begin{aligned}
    \Psi_n^R(\bm q,\omega)=&i\frac{1}{N}\sum_{\bm k,\bm k'}\int \frac{d\omega_1}{2\pi}v_n(\bm k, \bm q)G_{\alpha\alpha}(\bm k+\bm q,\omega+\omega_1)\\
    &\times G_{\beta\beta}(-\bm k,-\omega_1)M_{\bm k'}(\bm q)\Lambda_{\bm k\bm k'}(\bm q, \omega,\omega_1).
\end{aligned}
\end{equation}
The vector $\hat{\Psi}^R(\bm q,\omega)$ can be written as
\begin{equation}
    \hat{\Psi}^R(\bm q,\omega)=\hat{\Phi}^R(\bm q,\omega)+\hat{\chi}(\bm q,\omega)\hat{\Gamma}(\bm q)\hat{\Psi}^R(\bm q,\omega),
\end{equation}
where
\begin{equation}
\chi_{mn}(\bm q,\omega)
=
\frac{1}{N}
\sum_{\bm k}
v_m(\bm k, \bm q)
v_n(\bm k, \bm q)
\Pi_0(\bm q,\omega;\bm k).
\label{eq:chi_app}
\end{equation}
Finally, we obtain
\begin{equation}
G_{lad}(\bm q,\omega)=(\hat\Phi^L)^{T}
\hat\Gamma(\bm q)
\left[
\hat 1
-
\hat \chi(\bm q,\omega)
\hat\Gamma(\bm q)
\right]^{-1}
\hat \Phi^R(\bm q,\omega).
\end{equation}

\section{Separable Form of the Residual Magnon--Magnon Interaction Vertex}
\label{app:interaction_vertex}

The quartic magnon interaction entering the ladder approximation can be
written in the separable form
\begin{equation}
V_{\bm k\bm k'}(\bm q)
=
\frac{1}{N}
\sum_{mn}
v_m(\bm k, \bm q)
\Gamma_{mn}(\bm q)
v_n(\bm k', \bm q),
\end{equation}
where $N_c=18$ for the present model.

The basis function $v_n(\bm k, \bm q)$ is defined as
\begin{equation}
v(\bm{k},\bm q)
=
\begin{pmatrix}
u_{\bm{k+q}}u_{\bm{k}}\cos k_x \\
u_{\bm{k+q}}u_{\bm{k}}\sin k_x \\
u_{\bm{k+q}}u_{\bm{k}}\cos k_y \\
u_{\bm{k+q}}u_{\bm{k}}\sin k_y \\
u_{\bm{k+q}}v_{\bm{k}} \\
v_{\bm{k+q}}u_{\bm{k}} \\
v_{\bm{k+q}}v_{\bm{k}}\cos k_x \\
v_{\bm{k+q}}v_{\bm{k}}\sin k_x \\
v_{\bm{k+q}}v_{\bm{k}}\cos k_y \\
v_{\bm{k+q}}v_{\bm{k}}\sin k_y \\
u_{\bm{k+q}}v_{\bm{k}}\cos k_x\cos k_y \\
u_{\bm{k+q}}v_{\bm{k}}\sin k_x\cos k_y \\
u_{\bm{k+q}}v_{\bm{k}}\cos k_x\sin k_y \\
u_{\bm{k+q}}v_{\bm{k}}\sin k_x\sin k_y \\
v_{\bm{k+q}}u_{\bm{k}}\cos k_x\cos k_y \\
v_{\bm{k+q}}u_{\bm{k}}\sin k_x\cos k_y \\
v_{\bm{k+q}}u_{\bm{k}}\cos k_x\sin k_y \\
v_{\bm{k+q}}u_{\bm{k}}\sin k_x\sin k_y
\end{pmatrix}.
\end{equation}
For the conventional AFM, the interaction matrix $\Gamma_{mn}(\bm q)$ can be organized into the block form
\begin{equation}
\Gamma^{\rm AFM}=
\begin{pmatrix}
\Gamma_{uu} & \Gamma_{um} & 0 & 0\\
\Gamma_{um}^{T} & \Gamma_{mm} & \Gamma_{mv} & \Gamma_{md}\\
0 & \Gamma_{mv}^{T} & \Gamma_{vv} & 0\\
0 & \Gamma_{md}^{T} & 0 & \Gamma_{dd}
\end{pmatrix},
\end{equation}
where the individual blocks are
\begin{equation}
\Gamma_{uu}
=\Gamma_{vv}=
-2J_1 I_4,
~
\Gamma_{dd}
=
4J_2 I_8,
\end{equation}

\begin{equation}
\Gamma_{um}
=
\begin{pmatrix}
-J_1 & -J_1c_x\\
0 & J_1s_x\\
-J_1 & -J_1c_y\\
0 & J_1s_y
\end{pmatrix},
\end{equation}

\begin{equation}
\Gamma_{mv}
=
\begin{pmatrix}
-J_1c_x & J_1s_x &
-J_1c_y & J_1s_y\\
-J_1 & 0 &
-J_1 & 0
\end{pmatrix},
\end{equation}

\begin{equation}
\Gamma_{mm}
=
\begin{pmatrix}
C_1 & C_2\\
C_2 & C_1
\end{pmatrix},
\end{equation}

\begin{equation}
\Gamma_{md}
=
\begin{pmatrix}
C_3 & C_4 & C_5 & C_6 &
0 & 0 & 0 & 0\\
0 & 0 & 0 & 0 &
C_3 & C_4 & C_5 & C_6
\end{pmatrix}.
\end{equation}

The coefficients are defined by
\begin{align}
c_x&=\cos q_x,
&
s_x&=\sin q_x,
\nonumber\\
c_y&=\cos q_y,
&
s_y&=\sin q_y,
\\
C_1&=4J_2c_xc_y,
&
C_2&=-2J_1(c_x+c_y),
\nonumber\\
C_3&=-2J_2(1+c_xc_y),
&
C_4&=2J_2s_xc_y,
\nonumber\\
C_5&=2J_2c_xs_y,
&
C_6&=-2J_2s_xs_y.
\end{align}

For the AM $J_1$-$J_{2a}$-$J_{2b}$ model, only the blocks $\Gamma_{mm}$, $\Gamma_{md}$ and $\Gamma_{dd}$ are modified. The interaction matrix is symmetric, $\Gamma_{mn}=\Gamma_{nm}$, and we only list only the elements with $m\leq n$. The following matrix elements are replaced relative to the conventional AFM case:
\begin{equation}
\Gamma_{5,5}
=
\Gamma^{\rm AFM}_{5,5}
-
4\delta J_2\,s_xs_y,
\qquad
\Gamma_{6,6}
=
\Gamma^{\rm AFM}_{6,6}
+
4\delta J_2\,s_xs_y,
\end{equation}

\begin{equation}
\Gamma_{11,14}
=
\Gamma_{16,17}
=
-\Gamma_{12,13}
=
-\Gamma_{15,18}
=
-4\delta J_2,
\end{equation}

\begin{equation}
\begin{aligned}
\Gamma_{5,11}
&=
\Gamma^{\rm AFM}_{5,11}
+
2\delta J_2\,s_xs_y,\\
\Gamma_{5,12}
&=
\Gamma^{\rm AFM}_{5,12}
+
2\delta J_2\,c_xs_y,\\
\Gamma_{5,13}
&=
\Gamma^{\rm AFM}_{5,13}
+
2\delta J_2\,s_xc_y,\\
\Gamma_{5,14}
&=
\Gamma^{\rm AFM}_{5,14}
+
2\delta J_2\left(c_xc_y+1\right),\\
\Gamma_{6,15}
&=
\Gamma^{\rm AFM}_{6,15}
-
2\delta J_2\,s_xs_y,\\
\Gamma_{6,16}
&=
\Gamma^{\rm AFM}_{6,16}
-
2\delta J_2\,c_xs_y,\\
\Gamma_{6,17}
&=
\Gamma^{\rm AFM}_{6,17}
-
2\delta J_2\,s_xc_y,\\
\Gamma_{6,18}
&=
\Gamma^{\rm AFM}_{6,18}
-
2\delta J_2\left(c_xc_y+1\right).
\end{aligned}
\end{equation}

\section{Symmetry of the two-magnon ladder kernel at
\texorpdfstring{$\bm K$}{K}}
\label{app:Ksymmetry}
In this Appendix, we derive and prove the symmetry relations of the ladder kernel $\hat{A}=1-\hat\chi\hat\Gamma$ at $\bm K=(\pi/2,\pi/2)$, which lead to the twofold degeneracy of the low-energy resonance in the conventional antiferromagnetic phase and its splitting in the altermagnetic phase. 

\subsection{Matrix representation of momentum substitutions}
At $\bm K$, we consider two momentum substitutions 
\begin{equation}
    T:\bm k\rightarrow-\bm k-\bm K,
    \qquad
    R:\bm k\rightarrow-\bm k,
    \label{eq:app_TR}
\end{equation}
which lead to the transformations of Bogoliubov coefficients, 
\begin{equation}
\begin{aligned}
    T:&u_{\bm k}\rightarrow u_{\bm k+\bm K},~u_{\bm k+\bm K}\rightarrow u_{\bm k},\\
    &v_{\bm k}\rightarrow v_{\bm k+\bm K},~v_{\bm k+\bm K}\rightarrow v_{\bm k},
\end{aligned}
\end{equation}
\begin{equation}
\begin{aligned}
    R:&u_{\bm k}\rightarrow u_{\bm k},~u_{\bm k+\bm K}\rightarrow u_{\bm k+\bm K},\\
    &v_{\bm k}\rightarrow v_{\bm k},~v_{\bm k+\bm K}\rightarrow -v_{\bm k+\bm K},
\end{aligned}
\end{equation}
These relations allow the two momentum substitutions to
be represented directly in the separable channel space as
\begin{equation}
    \bm v(T\bm k,\bm K)
    =
    \hat S_T\bm v(\bm k,\bm K),
    \qquad
    \bm v(R\bm k,\bm K)
    =
    \hat R_D\bm v(\bm k,\bm K),
    \label{eq:app_channeltransform}
\end{equation}
where 
\begin{equation}
    \hat S_T=(-\hat M_2)\oplus (-\hat M_2)\oplus \hat M_2\oplus (-\hat M_2)\oplus (-\hat M_2)\oplus \hat M_8,
\end{equation}
\begin{equation}
\begin{aligned}
\hat R_D = {\rm diag}
[&1,-1,1,-1,1,-1,-1,1,-1,1,\\
&1,-1,-1,1,-1,1,1,-1],
\end{aligned}
\label{eq:app_RD}
\end{equation}
with $(\hat M_n)_{ij} = \delta_{i,n+1-j}$.
Both are orthogonal matrices,
\begin{equation}
    \hat S_T^2=\hat R_D^2=\hat 1.
    \label{eq:app_involutions}
\end{equation}
Interchanging the order of acting $T$ and $R$ gives
\begin{equation}
    TR\,\bm k=\bm k-\bm K,
    \qquad
    RT\,\bm k=\bm k+\bm K,
\end{equation}
and the two resulting momenta differ by $\bm M=(\pi,\pi)$. Using the relations,
\begin{equation}
    u_{\bm{k+M}}=u_{\bm k},
    \qquad
    v_{\bm{k+M}}=-v_{\bm k},
    \label{eq:app_uvM}
\end{equation}
we obtain
\begin{equation}
    \bm v(\bm{k+M},\bm K)
    =
    -\bm v(\bm k,\bm K),
\end{equation}
which gives the anticommutation relation between $T$ and $R$,
\begin{equation}
\{\hat S_T,\hat R_D\}=0.
\label{eq:app_anticommute}
\end{equation}

\subsection{Twofold degeneracy for \texorpdfstring{$\delta J_2=0$}
{delta J2=0}}
In the conventional antiferromagnetic phase ($\delta J_2=0$), we label the matrices in ladder kernel $\hat{A}=1-\hat\chi\hat\Gamma$ at $\bm K=(\pi/2,\pi/2)$ by $\hat{A}_0$, $\hat{\chi}_0$ and $\hat{\Gamma}_0$. Beginning with the two-magnon energy, 
\begin{equation}
    E_2(\bm k,\bm K)=\bar{\omega}(\bm k+\bm K)+\bar{\omega}(\bm k),
\end{equation}
both transformation $T$ and $R$ leave $E_2(\bm k,\bm K)$ unchanged,
\begin{equation}
\begin{aligned}
    E_2(T\bm k,\bm K)&=\bar{\omega}(-\bm k)+\bar{\omega}(\bm k+K)=E_2(\bm k,\bm K),\\
    E_2(R\bm k,\bm K)&=\bar{\omega}( -\bm k+\bm K)+\bar{\omega}(-\bm k)=E_2(\bm k,\bm K),
\end{aligned}
\end{equation}
thus the bare two-magnon propagator $\Pi_0$ satisfies
\begin{equation}
\begin{aligned}
    \Pi_0(T\bm k,\bm K,\omega;0)&=\Pi_0(\bm k,\bm K,\omega;0),\\
    \Pi_0(R\bm k,\bm K,\omega;0)&=\Pi_0(\bm k,\bm K,\omega;0).
\end{aligned}
\end{equation}
Thus, the matrix $\hat{\chi}$
\begin{equation}
    \chi_{mn}=\frac{1}{N}\sum_{\bm k}v_m(\bm k,\bm K)v_n(\bm k,\bm K)\Pi_0(\bm k,\bm K,\omega)
    \label{eq:app_chi}
\end{equation}
keeps unchanged under these two momentum substitutions, because it is just the changing of summation variable,
\begin{equation}
    \hat S_T\hat\chi_0\hat S_T
    =
    \hat\chi_0,
    \qquad
    \hat R_D\hat\chi_0\hat R_D
    =
    \hat\chi_0.
    \label{eq:app_chisym0}
\end{equation}
At $\bm q=\bm K$, the residual magnon-magnon interaction obeys the same transformations, 
\begin{equation}
    \hat S_T\hat\Gamma_0\hat S_T
    =
    \hat\Gamma_0,
    \qquad
    \hat R_D\hat\Gamma_0\hat R_D
    =
    \hat\Gamma_0,
    \label{eq:app_Gammasym0}
\end{equation}
Thus, we obtain
\begin{equation}
    [\hat S_T,\hat A_0]=[\hat R_D,\hat A_0]=0.
\label{eq:app_Acommute0}
\end{equation}

The twofold degeneracy now follows immediately from
Eq.~(\ref{eq:app_anticommute}). Let
\begin{equation}
    \hat A_0 w=a\,w,
    \qquad
    \hat R_Dw=r\,w,
    \qquad r=\pm1.
\end{equation}
Because $\hat S_T$ commutes with $\hat A_0$,
\begin{equation}
    \hat A_0(\hat S_Tw)
    =
    a\,\hat S_Tw.
\end{equation}
At the same time, the anticommutation relation gives
\begin{equation}
    \hat R_D(\hat S_Tw)
    =
    -\hat S_T\hat R_Dw
    =
    -r\,\hat S_Tw.
\end{equation}
Thus $w$ and $\hat S_Tw$ have the same eigenvalue of $\hat A_0$ but
opposite $\hat R_D$ parity, and are therefore linearly independent. The
spectrum of $\hat A_0(\bm K,\omega,0)$ is consequently twofold
degenerate at every frequency. In particular, the near-null modes
associated with the low-energy resonance form a degenerate pair in the
conventional antiferromagnetic phase.

\subsection{Finite altermagnetic exchange anisotropy}
We next consider the altermagnetic phase with finite $\delta J_2$. The two-magnon energy at $\bm K$ is 
\begin{equation}
    E_2(\bm k,\bm K;\delta J_2)=
    \omega_+(\bm{k+K};\delta J_2)+\omega_-(-\bm k;\delta J_2).
\label{eq:app_E2}
\end{equation}

Under the momentum substitution $T$, the two magnons are exchanged and the sign of the altermagnetic splitting is reversed. Consequently, we have
\begin{equation}
\begin{aligned}
   E_2(T\bm k,\bm K;\delta J_2)=&\omega_+(-\bm{k};\delta J_2)+\omega_-(\bm{k+K};\delta J_2)\\
   =&\bar{\omega}(-\bm k)+\bar{\omega}(\bm{k+K})\\
   &+\Delta(-\bm k)-\Delta(\bm{k+K})\\
   =&E_2(\bm k,\bm K;-\delta J_2),
\label{eq:app_E2T}
\end{aligned}
\end{equation}
and therefore
\begin{equation}
    \Pi_0(T\bm k,\bm K,\omega;\delta J_2)
    =
    \Pi_0(\bm k,\bm K,\omega;-\delta J_2),
    \label{eq:app_PiT}
\end{equation}
which gives
\begin{equation}
    \hat S_T
    \hat\chi(\omega,\delta J_2)
    \hat S_T=
    \hat\chi(\omega,-\delta J_2).
\label{eq:app_chiTdelta}
\end{equation}
For the residual interaction matrix $\hat\Gamma=\hat\Gamma_0+\delta J_2\hat\Gamma_1$, we have
\begin{equation}
    \hat S_T\hat\Gamma_0\hat S_T=\hat\Gamma_0,
    \qquad
    \hat S_T\hat\Gamma_1\hat S_T=-\hat\Gamma_1,
\end{equation}
which is equivalent to
\begin{equation}
    \hat S_T
    \hat\Gamma(\delta J_2)
    \hat S_T
    =
    \hat\Gamma(-\delta J_2).
    \label{eq:app_GammaTdelta}
\end{equation}
Combine Eq.~(\ref{eq:app_chiTdelta}) and Eq.~(\ref{eq:app_GammaTdelta}), we obtain
\begin{equation}
\hat S_T
    \hat A(\bm K,\omega,\delta J_2)
    \hat S_T
    =
    \hat A(\bm K,\omega,-\delta J_2).
\label{eq:app_Adelta}
\end{equation}
For the other momentum substitution $R$, we have
\begin{equation}
\begin{aligned}
   E_2(R\bm k,\bm K;\delta J_2)&=\omega_+(-\bm k+\bm{K};\delta J_2)+\omega_-(\bm k;\delta J_2)\\
   &=E_2(\bm k,\bm K;\delta J_2),
\end{aligned}
\end{equation}
where the two-magnon pair energy remains invariant. Thus, the matrix $\hat{\chi}$ is unchanged under $R$,
\begin{equation}
    \hat R_D
    \hat\chi(\omega,\delta J_2)
    \hat R_D
    =
    \hat\chi(\omega,\delta J_2).
\end{equation}
The interaction matrix $\hat\Gamma$ is also even under $\hat R_D$ at $\bm K$,
\begin{equation}
    \hat R_D
    \hat\Gamma(\delta J_2)
    \hat R_D
    =
    \hat\Gamma(\delta J_2).
\end{equation}
Thus, $\hat{R}_D$ always commutes with $\hat A(\bm K,\omega,\delta J_2)$,
\begin{equation}
    [\hat R_D,
    \hat A(\bm K,\omega,\delta J_2)]=0.
\label{eq:app_RDdelta}
\end{equation}
Therefore, the resonance components at finite $\delta J_2$ are still labeled by their $\hat R_D$ parity,
\begin{equation}
    \hat R_Dw_r=r\,w_r,
    \qquad
    r=\pm1.
\end{equation}
Let $\Omega_{\bm K,r}(\delta J_2)$ denote the resonance position in
the $r$ sector, with the corresponding near-null mode satisfying
\begin{equation}
    \hat A\bigl(\bm K,\Omega_{\bm K,r}(\delta J_2),\delta J_2\bigr)
    w_r
    \simeq 0.
\end{equation}
Under $\hat S_T$, we obtain
\begin{equation}
\hat A\bigl(\bm K,\Omega_{\bm K,r}(\delta J_2),-\delta J_2\bigr)
\hat S_T w_r\simeq 0.
\end{equation}
Since $\{\hat S_T,\hat R_D\}=0$, the transformed mode has the opposite
parity,
\begin{equation}
    \hat R_D(\hat S_Tw_r)
    =
    -r\,\hat S_Tw_r.
\end{equation}
Thus, $\hat S_T$ maps the $r$ resonance sector at $\delta J_2$ onto
the $-r$ sector at $-\delta J_2$ at the same frequency, giving
\begin{equation}
    \Omega_{\bm K,r}(\delta J_2)
    =
    \Omega_{\bm K,-r}(-\delta J_2),
    \label{eq:app_Omegasym}
\end{equation}
which directly leads to the conclusion that the signed splitting is odd in
$\delta J_2$, whereas the midpoint is even,
\begin{equation}
\begin{aligned}
    \Omega_{\bm K,+}-\Omega_{\bm K,-}
    &=
    b_1\delta J_2+b_3\delta J_2^3+\cdots,\\
    \frac{\Omega_{\bm K,+}+\Omega_{\bm K,-}}{2}
    &=
    \Omega_0+a_2\delta J_2^2+\cdots.
\label{eq:app_odd_even}
\end{aligned}
\end{equation}
The leading allowed contribution to the resonance splitting is therefore linear in $\delta J_2$, whereas the midpoint has no linear correction, consistent with the numerical results discussed in the main text.

\subsection{Numerical resonance positions and spectral peaks}

We finally compare the resonance positions obtained directly from the ladder kernel with the peak positions of the full two-magnon response at $\bm K=(\pi/2,\pi/2)$. For each $\hat R_D$ parity sector, we identify $\Omega_{\bm K,r}$ from the minimum modulus of the eigenvalue $a_r(\omega)$ of $\hat A$,
\begin{equation}
    \Omega_{\bm K,r}
    =
    \mathop{\rm argmin}_{\omega}
    |a_r(\omega)|,
    \qquad
    r=\pm1.
    \label{eq:app_Omega_num}
\end{equation}
The resulting resonance positions are summarized in Table~\ref{table:K_resonance_positions}. For $\delta J_2=0$, the two parity sectors give the same resonance position, $\Omega_{\bm K,+}=\Omega_{\bm K,-}\simeq2.110J_1$, consistent
with the twofold degeneracy derived above. With increasing
$\delta J_2$, the $r=+1$ resonance shifts toward higher energy while the $r=-1$ resonance shifts toward lower energy. 
The minimum values of $|a_r|$ remain finite because the resonance lies inside the two-magnon continuum.
\begin{table}
\centering
\begin{tabular}{|c|c|c|c|c|c|}
\hline
$\delta J_2$
& $\Omega_{\bm K,-}$
& $\Omega_{\bm K,+}$
& $\omega_{\bm K}^{\rm lower}$
& $\omega_{\bm K}^{\rm upper}$
& mid \\
\hline
0.00 & 2.1100 & 2.1100 & 2.1120 & 2.1120 & 2.1100 \\
0.04 & 2.0633 & 2.1535 & 2.0674 & 2.1544 & 2.1084 \\
0.08 & 2.0162 & 2.1944 & 2.0194 & 2.1950 & 2.1053 \\
0.12 & 1.9682 & 2.2312 & 1.9703 & 2.2320 & 2.0997 \\
0.16 & 1.9186 & 2.2647 & 1.9196 & 2.2654 & 2.0916 \\
\hline
\end{tabular}
\caption{The resonance positions obtained from the ladder
kernel and the corresponding peak positions of the full two-magnon
intensity at $\bm K=(\pi/2,\pi/2)$. The midpoint shown here is obtained
from the kernel resonance positions,
$(\Omega_{\bm K,+}+\Omega_{\bm K,-})/2$.}
\label{table:K_resonance_positions}
\end{table}

The resonance splitting is well described by a linear dependence over
the parameter range studied,
\begin{equation}
    \Omega_{\bm K,+}-\Omega_{\bm K,-}
    \simeq 2.18\,\delta J_2,
    \label{eq:app_Omega_split_fit}
\end{equation}
while the midpoint follows
\begin{equation}
    \frac{\Omega_{\bm K,+}+\Omega_{\bm K,-}}{2}
    \simeq
    2.110J_1
    -0.72\delta J_2^2.
    \label{eq:app_Omega_mid_fit}
\end{equation}
These results are consistent with Eq.~(\ref{eq:app_odd_even}).

We next compare these intrinsic resonance positions with the peaks of the full two-magnon intensity. For $\delta J_2>0$, the $r=\pm1$ resonance components closely follow the upper and lower spectral peaks. The peak separation extracted from the intensity is
\begin{equation}
    \omega_{\bm K}^{\rm upper}
    -
    \omega_{\bm K}^{\rm lower}
    \simeq
    2.17\,\delta J_2,
    \label{eq:app_peak_split_fit}
\end{equation}
which is in excellent agreement with the slope $2.18$ obtained directly from the ladder kernel. The midpoint of the two spectral peaks follows
\begin{equation}
    \frac{
    \omega_{\bm K}^{\rm upper}
    +
    \omega_{\bm K}^{\rm lower}}{2}
    \simeq
    2.112J_1
    -0.76\delta J_2^2,
    \label{eq:app_peak_mid_fit}
\end{equation}
which also closely agrees with the kernel result. Thus, the two peaks
observed in the full response directly track the splitting of the
two resonance components of the interacting two-magnon
kernel.

\bibliographystyle{unsrt}
\bibliography{refs}

\end{document}